\documentclass{article}
\def\Th{\Theta}
\def\be{{\bf e}}
\def\bE{{\bf E}}
\def\bH{{\bf H}}

\def\ud1{\dot{u}_{1}}
\def\bs{{\mbox{\boldmath{$\sigma$}}}}
\def\bw{{\mbox{\boldmath{$\omega$}}}}
\def\beq{\begin{equation}}
\def\eeq{\end{equation}}
\def\brr{\begin{array}}
\def\err{\end{array}}
\def\bea{\begin{eqnarray}}
\def\eea{\end{eqnarray}}
\def\bean{\begin{eqnarray*}}
\def\eean{\end{eqnarray*}}

\newcommand{\sfrac}[2]{{\textstyle{#1\over#2}}}
\newtheorem{defi}{Definition}[section]
\def\case#1/#2{\textstyle\frac{#1}{#2}}

\def\be{{\bf e}}

\def\hsh{\hspace*{.5cm}}

\def\bi{\bibitem}
\def\ct{\cite}
\def\l{\label}
\def\r{\ref}
\def\cqg{{\em Class. Quantum Grav.\/} }
\def\grg{{\em Gen. Rel. Grav.\/} }
\def\prd{{\em Phys. Rev.\/} D }
\def\prl{{\em Phys. Rev. Lett.\/} }
\def\jmp{{\em J. Math. Phys.\/} }
\def\cmp{{\em Commun. Math. Phys.\/} }
\title{{\sc Partially locally rotationally symmetric perfect fluid
cosmologies}}
\author{{\sc Nazeem Mustapha\thanks{E-mail: {\tt
nazeem@dyson.mth.uct.ac.za}}, George F R Ellis\thanks{E-mail: {\tt
ellis@maths.uct.ac.za}}}, \\ {\sc Henk van Elst\thanks{E-mail:
{\tt henk@gmunu.mth.uct.ac.za}} \ \& Mattias
Marklund\thanks{E-mail: {\tt mattias.marklund@sto.foa.se}}
\thanks{Presently at: National Defence Research Establishment,
SE-172 90 Stockholm, Sweden}}\\ {\small\em Cosmology Group,
Department of Mathematics and Applied Mathematics, University of
Cape Town}\\ {\small\em Rondebosch 7701, Cape Town, South Africa}}
\date{\normalsize{May 16, 1999}}
\begin{document}
\sloppy
\maketitle
\begin{abstract}
We show that there are no new consistent cosmological perfect fluid
solutions when in an open neighbourhood ${\cal U}$ of an event the
fluid kinematical variables and the electric and magnetic Weyl
curvature are all assumed rotationally symmetric about a common
spatial axis, specialising the Weyl curvature tensor to algebraic
Petrov type D. The consistent solutions of this kind are either
locally rotationally symmetric, or are subcases of the Szekeres
dust models. Parts of our results require the assumption of a
barotropic equation of state. Additionally we demonstrate that
local rotational symmetry of perfect fluid cosmologies follows from
rotational symmetry of the Riemann curvature tensor and of its
covariant derivatives only up to second order, thus strengthening a
previous result.
\begin{flushleft}
Preprint: uct--cosmology--99/09
\end{flushleft}
%

\end{abstract}
\centerline{\bigskip\noindent PACS number(s): 04.20.-q, 98.80.Hw, 
04.20.Jb}

\section{Introduction}
Our long-term aim is to determine all perfect fluid cosmologies
which cannot be invariantly defined by the existence of a unique
eigentetrad for the {\em fluid shear\/} tensor
$\mbox{\boldmath{$\sigma$}}$. This consideration is of relevance to
the equivalence problem of spacetimes
\ct{bi:C46,bi:K80} and its application to relativistic
cosmology. The fluid shear plays a central role in the dynamics of
a generic cosmology \ct{bi:E71}. If it vanishes, the resultant
consistent models are highly special and well-studied for perfect
fluid matter sources \ct{bi:CW83}. In general,
${\mbox{\boldmath{$\sigma$}}}$ has two distinct eigenvalues, which
may serve to invariantly classify cosmologies on the basis of the
related eigentetrad. When these eigenvalues coincide (degenerate),
but with $\mbox{\boldmath{$\sigma$}} \neq 0$, the
$\mbox{\boldmath{$\sigma$}}$--eigentetrad is no longer uniquely
defined and the fluid shear tensor thus picks out a distinct
spatial direction. It is the determination of all cosmological
perfect fluid solutions exhibiting this feature that interests us
in the long run.

Key symmetries of spacetimes are their continuous isotropies, and
cosmologies are either {\em isotropic\/} (and then have a
Robertson--Walker metric), {\em locally rotationally symmetric}
(`LRS') (and then are all known up to the {\em form\/} of their
metric), or are {\em anisotropic\/} (see, e.g., \ct{bi:E67} and
\ct{bi:EvE99} for a discussion of these cases). In the case of LRS
cosmologies, there is at each event (relative to the family of
fundamental observers) precisely one preferred spatial direction,
and all physical properties and observations are invariant under
rotation about this direction. It follows that these spacetimes are
invariant under multiply transitive groups of isometries
\ct{bi:E67,bi:SE68,bi:vEE96}. The question that is interesting from
both the physical point of view, and in terms of determining the
equivalence of cosmologies, is {\em how weak\/} we can make the
assumptions of rotational symmetry and still determine explicitly
the family of cosmological spacetimes involved; in other words,
{\em how few\/} physical and geometrical quantities we can make
{\em rotationally symmetric\/}, where rotationally symmetric means
the quantity concerned is either isotropic, or invariant under
arbitrary rotations about a preferred spatial axis.

Central to the equivalence problem formalism are the components of
the spacetime Riemann curvature tensor ${\bf R}$ and its covariant
derivatives ${\bf \nabla\dots\nabla R}$ in a standard tetrad. In
\ct{bi:E67} it was shown that a spacetime will be LRS if all
tensors algebraically defined by the spacetime Ricci curvature
tensor and their covariant derivatives up to {\em third order\/}
are rotationally symmetric (note that the fluid 4-velocity field
${\bf u}$ is algebraically determined by the Ricci curvature
tensor, through the Einstein field equations). The ultimate aim is
to weaken this assumption by considering perfect fluid cosmologies
in which {\em only\/} the fluid shear tensor
$\mbox{\boldmath{$\sigma$}}$ (and not its covariant derivatives)
has this symmetry. All cosmologies that do {\em not\/} satisfy this
restriction can be invariantly defined through tensor components
relative to the unique
$\mbox{\boldmath{$\sigma$}}$--eigentetrad. In the present work we
set off to consider a {\em subcase\/} of this more general project
in the context of an approach that is applicable to investigating
systematically all cases with rotationally symmetric
$\mbox{\boldmath{$\sigma$}}$. In detail, we consider perfect fluid
cosmologies in which all {\em fluid kinematical\/} and {\em Weyl
curvature variables\/} are {\em rotationally symmetric about the
same spatial axis\/} (the Ricci curvature tensor components are
automatically so, because of the perfect fluid assumption), thus
restricting the cosmologies to algebraic Petrov type D and simpler
cases. We make no similar assumption about the covariant
derivatives of these variables. We find all perfect fluid
cosmologies satisfying this restriction.

\subsection{Assumptions}
We describe a spacetime by a pseudo-Riemannian manifold ${\cal M}$
with a rank two symmetric metric tensor field ${\bf g}$. It is
convenient to decompose the Riemann curvature tensor ${\bf R}$ into
the completely tracefree Weyl conformal curvature tensor
$C_{abcd}$, the Ricci curvature tensor $R_{ab} := R^{c}{}_{acb} =
R_{(ab)}$ and the Ricci curvature scalar $R := g^{ab}R_{ab}$
according to\footnote{When referring to tetrad components in the
spacetime, we use Latin indices from the first half of the
alphabet: $a, \,b, \,c \in \{\,0, \,1, \,2, \,3\,\}$; in the
spatial sections we use Greek letters: $\alpha, \,\beta, \,\gamma
\in \{\,1, \,2, \,3\,\}$. For the local coordinate spacetime
description we use Latin indices from the second half of the
alphabet: $i, \,j, \,k \in \{\,0, \,1, \,2, \,3\,\}$. Round
brackets denote symmetrised indices and square brackets denote
skew-symmetrised indices.}
\beq
\l{eq:decomposedWeyl}
R_{abcd} = C_{abcd} + R_{a[c}\,g_{d]b} - R_{b[c}\,g_{d]a}
- \sfrac{1}{3}\,R\,g_{a[c}\,g_{d]b} \ .
\eeq
We will mean by a {\em cosmology\/} a spacetime which satisfies
all of the following requirements \ct{bi:E61,bi:E67,bi:EvE99}:
\begin{itemize}

\item  
It is a self-consistent solution of the {\em Einstein field
equations\/} (`EFE'), relating matter source fields which are
represented by an energy--momentum--stress tensor $T_{ab}$ to the
Ricci curvature tensor $R_{ab}$ and its trace $R$
as\footnote{Throughout this work we will employ geometrised units
characterised by $c = 1 = 8\pi G/c^{2}$, and we set the
cosmological constant $\Lambda$ equal to zero (the latter implies
no loss of generality, as $\Lambda$ can be effectively included by
suitably redefining $T_{ab}$).}
\beq
\l{eq:EFE}
R_{ab}-\sfrac{1}{2}\,R\,g_{ab} = T_{ab} \ .
\eeq

\item 
It is filled with matter energy of some sort which we can represent
as a {\em perfect fluid\/} with fundamental {\em unit 4-velocity\/}
field ${\bf u}$ ($u_au^a = -1$). The energy--momentum--stress
tensor $T_{ab}$ is then given by
\beq
\l{eq:emom}
T_{ab} = \mu\,u_a\,u_b + p\,h_{ab} = T_{(ab)} \ ,
\eeq
where $h_{ab} := g_{ab} + u_au_b$ denotes the tensor projecting
into the rest 3-spaces of observers moving with 4-velocity ${\bf
u}$, and $\mu := T_{ab}u^au^b$ and $p :=
\sfrac{1}{3}\,T_{ab}h^{ab}$, respectively, are the total energy
density and isotropic pressure measured in these rest 3-spaces.

\item
Additionally we assume
\[
(\mu+p) > 0 \ ,
\]
so that by (\r{eq:EFE}), together with (\r{eq:emom}), ${\bf u}$
uniquely defines the {\em timelike\/} eigenvector of $R_{ab}$.

\item  
Observations have shown that the Universe is expanding. We will
take this to mean that the (isotropic) expansion of ${\bf u}$,
described by the {\em fluid expansion\/} scalar $\Th$, is positive:
\[
\Th > 0 \ .
\]
\end{itemize}
The Weyl conformal curvature tensor may be decomposed relative to
${\bf u}$ into its symmetric tracefree `electric' and `magnetic'
tensor parts, $\bE$ and $\bH$, respectively, according to
\ct{bi:E61, bi:E71}
\bea
\l{eq:E_ab}
E_{ab} & := & C_{cedf}\,h^{c}{}_{a}\,u^{e}\,h^{d}{}_{b}\,u^{f}
\ = \ E_{(ab)} \\
\l{eq:H_ab}
H_{ab} & := & (-\,\sfrac{1}{2}\,\eta_{cegh}\,
C^{gh}{}_{df})\,h^{c}{}_{a}\,u^{e}\,h^{d}{}_{b}\,u^{f}
\ = \ H_{(ab)} \ ,
\eea
with the completely skew spacetime permutation tensor $\eta_{abcd}$
specified by $\eta_{abcd} = \eta_{[abcd]}$ and $\eta _{0123} = 1$,
$\eta^{0123} = -\,1$. The second Bianchi identities differentially
relate components of the Riemann curvature tensor:
\beq
\l{eq:Bianchi2}
\nabla _{[a}R_{bc]de} = 0 \ .
\eeq
As well as entailing the matter equations of motion
$\nabla_{b}T^{ab} = 0 = \nabla_{b}(R^{ab}-\sfrac{1}{2}\,
R\,g^{ab})$, this relation provides evolution and constraint
equations for ${\bf E}$ and ${\bf H}$ when employing the $1+3$
decompositions (\r{eq:E_ab}) and (\r{eq:H_ab})
\ct{bi:E71,bi:EvE99}.

\subsection{The problem: symmetry assumptions}
The question we answer here is: `Under what conditions are all of
the fluid acceleration, vorticity and shear and the spacetime
electric and magnetic Weyl curvatures {\em either\/} rotationally
symmetric about the same spatial axis {\em or\/} isotropic, but the
spacetime itself is not?'. A dynamical tensor field is rotationally
symmetric if there is a degeneracy in that tensor quantity in terms
of its eigenvalues, but it is not isotropic (not all eigenvalues
are zero). The justification for this terminology is based on the
fact that a spacetime is LRS if all tensor quantities, as well as
their covariant derivatives, are either isotropic or rotationally
symmetric about the same spatial axis, with {\em at least one not
being isotropic\/}
\ct{bi:E67,bi:vEE96}. Our ultimate aim is to be able to uniquely
classify all perfect fluid cosmologies which have {\em degenerate
fluid shear eigenvalues\/}. To this end, we employ the methods
developed in \ct{bi:E67} which introduce as a reference framework
an orthonormal tetrad (`ONT') field $\{\,\be_{a}\,\}$ where the
timelike member $\be_{0}$ is identified with the fluid 4-velocity
field ${\bf u}$ (`$1+3$ decomposition'). For the spatial rotation
coefficients we use the notation suggested in
\ct{bi:EM69,bi:M73,bi:vEU97} and utilised to great effect by
Wainwright and collaborators (see
\ct{bi:WE97} for a survey). We have thus written out for a perfect
fluid matter source the {\em evolution\/} and {\em constraint
equations\/}\footnote{When $\be_{0} = {\bf {u}}$ has {\em
non-vanishing\/} vorticity this terminology is doubtful because
then the local rest 3-spaces of neighbouring fundamental observers
do {\em not\/} mesh to form everywhere spacelike 3-surfaces.} for
the tetrad commutation functions $\mbox{\boldmath{$\gamma$}}$ as
well as for $\bE$ and $\bH$, which are obtained from the Jacobi,
Ricci and second Bianchi identities, and specialised to a
rotationally symmetric fluid shear tensor
$\mbox{\boldmath{$\sigma$}}$. These are given in full in appendix
\r{ap:fullset}.



\section{PLRS perfect fluid cosmologies}
\l{se:sEH_LRS}
The effect of restricting the geometry of a spacetime is generally
expressed by setting certain geometrically defined tensor
components to {\em zero\/}. The evolution equations for these
quantities now become {\em new constraint equations\/}. These
constraint equations must be preserved along the fluid flow lines
${\bf u}$ to be consistent. In general this does not happen without
further constraints on the dynamics. Of course, the test for
preservation then has to be repeated for these new constraints. We
list all perfect fluid cosmologies that provide solutions to the
EFE which have the tensor fields ${\mbox{\boldmath{$\sigma$}}}$,
${\bf E}$ and ${\bf H}$ as well as the vector fields $\dot{{\bf
u}}$ and ${\mbox{\boldmath{$\omega$}}}$, {\em but not their
covariant derivatives\/}, all rotationally symmetric about the same
spatial axis. In our investigation we find that all consistent
cosmological solutions to the EFE satisfying this criterion, bar
one --- the Szekeres dust solutions ---, are in fact LRS spacetimes
in the definition of Ellis \ct{bi:E67} and Stewart and Ellis
\ct{bi:SE68}. We shall generally take $\mbox{\boldmath{$\sigma$}}
\neq 0$, as for $\mbox{\boldmath{$\sigma$}} = 0$ we will see that
the presently considered types of cosmologies typically specialise
to the well known spatially homogeneous and isotropic
Friedmann--Lema\^{\i}tre--Robertson--Walker (`FLRW') cases.
\begin{defi}
\l{def:1}
We assume that in an open neighbourhood ${\cal {U}}$ of an event on
an expanding perfect fluid spacetime manifold ${\cal M}$
\begin{eqnarray}
\l{eq:PLRS}
(\sigma_{22}-\sigma_{33})
= \sigma_{12} = \sigma_{23} = \sigma_{31} = 0 \ , & & \nonumber \\
(E_{22}-E_{33}) = E_{12} = E_{23} = E_{31} = 0 \ , & & \nonumber \\
(H_{22}-H_{33}) = H_{12} = H_{23} = H_{31} = 0 \ , & & \\
\dot{u}_{2} = \dot{u}_{3} = 0 \ , & & \nonumber \\
\omega_{2} = \omega_{3} = 0 \ . & & \nonumber
\end{eqnarray}
That is to say, we have set all of ${\mbox{\boldmath{$\sigma$}}}$,
${\bf E}$ and ${\bf H}$ simultaneously rotationally symmetric about
the same spatial axis. In addition, the vectors $\dot{{\bf u}}$ and
${\mbox{\boldmath{$\omega$}}}$ are aligned with this axis of
rotational symmetry. We call all cosmological spacetimes in which
there exists a tetrad $\{\,\be_{a}\,\}$ with respect to which these
restrictions are satisfied {\em partially locally rotationally
symmetric (`PLRS')\/} cosmologies. Those PLRS cosmologies that are
{\em not\/} LRS will be referred to as {\em strictly PLRS\/}.
\end{defi}

\noindent
{\em Remarks\/}:

(i) Using a ${\mbox{\boldmath{$\sigma$}}}$--eigentetrad
$\{\,\be_{a}\,\}$ here is equivalent to using an ${\bf E}$-- or
${\bf H}$--eigentetrad.

(ii) If either of the vector quantities $\dot{{\bf u}}$ or
${\mbox{\boldmath{$\omega$}}}$ do {\em not\/} point in the
invariant spatial direction defined by the degenerate tensor
quantities, then they define an additional invariant spatial
direction. The tetrad $\{\,\be_{a}\,\}$ may then be invariantly
defined by, e.g., identifying another of its members with this new
direction. And thus, for the situations we want to start from to be
PLRS, we must have that ${{\dot{u}_{2}}} = {{\dot{u}_{3}}} = 0$;
that is, $\dot{{\bf u}} \parallel \be_{1}$. Similarly, we must have
${{{\omega}_{2}}} = {{{\omega}_{3}}} = 0$, corresponding to
${\mbox{\boldmath{$\omega$}}} \parallel \be_{1}$. If these
conditions did not hold, this would exclude for non-zero $\dot{{\bf
u}}$ and ${\mbox{\boldmath{$\omega$}}}$ the possibility of the
spacetime being LRS and thus compromise the presently proposed
notion of partial symmetry.

(iii) The present setup arbitrarily adapts to the spatial ${\bf
e}_{1}$--axis. However, this is only a matter of convention and by
a cyclic permutation of indices $1 \rightarrow 2 \rightarrow 3
\rightarrow 1$ one can easily adapt to any of the other spatial
axes as well.
 
(iv) Our assumptions reduce the perfect fluid cosmologies we
consider in this article to algebraic Petrov type D; a
classification scheme for solutions with this property based on the
Newman--Penrose formalism has been suggested by Wainwright in
\ct{bi:W76}.  \\
 
With Definition \r{def:1} in place, the momentum conservation
equations for all PLRS cosmologies reduce in ${\cal U}$ to
(\r{eq:1b;b}) -- (\r{eq:3b;b}). Moreover, combining the
${\bH}$--constraint equations (\r{eq:R0231}) with (\r{eq:R0312})
and (\r{eq:R0331}) with (\r{eq:R0212}), we find that all PLRS
cosmologies require in ${\cal U}$ that
$$(n_{22}-n_{33})\,(\sigma_{11}^{2}+\sfrac{4}{9}\,
\omega_{1}^{2}) = n_{23}\,(\sigma_{11}^{2}+\sfrac{4}{9}\,
\omega_{1}^{2}) = 0;$$ which means, as we assume that 
${\mbox{\boldmath{$\sigma$}}} {\mbox{\boldmath{$\omega$}}} \neq 0$,
that\footnote{This restriction on $n_{\alpha\beta}$ suggests that
for PLRS cosmologies according to Definition \r{def:1} no
consistent solutions to the EFE exist that contain {\em
gravitational radiation\/}; it is the {\em transverse\/} (with
respect to the spatial ${\bf e}_{1}$--direction) components
$(n_{22}-n_{33})$, $n_{23}$, $(\sigma_{22}-\sigma_{33})$ and
$\sigma_{23}$ which typically form the {\em connection\/}
characteristic eigenfields associated with the local null cones
(cf. \ct{bi:vEE99}). If we relax the PLRS restrictions so that
${\bE}$ is not rotationally symmetric, we find that this feature
still holds.}
\beq
\l{eq:zerotrn}
(n_{22}-n_{33}) = n_{23} = 0 \ .
\eeq
The related evolution equations obtained from (\r{eq:Jac2013}) and
(\r{eq:Jac3012}) as well as (\r{eq:Jac2012}) and (\r{eq:Jac3013}),
respectively, now reduce in ${\cal U}$ to constraints on the
$\be_{2}$-- and $\be_{3}$--gradients of the Fermi-rotation
variables $\Omega_{2}$ and $\Omega_{3}$; namely
\bea
\l{eq:fomc1}
(\be_{2}+2\,n_{31})\,(\Omega_{2})
- (\be_{3}-2\,n_{12})\,(\Omega_{3}) & = & 0 \\
\l{eq:fomc2}
(\be_{2}+2\,n_{31})\,(\Omega_{3})
+ (\be_{3}-2\,n_{12})\,(\Omega_{2}) & = & 0 \ .
\eea

We proceed to check the consistency of the cosmological PLRS
subcase of the EFE by mainly computing the time evolution of all
new constraints. In the process of doing this we will fix the
remaining tetrad freedom and thus invariantly classify the
solutions. For each of the cases we consider below we will check
the transformation behaviour of the commutation functions and other
tensor quantities and use this to fix the freedom conveniently.

We now turn to the issue of choice of spatial triad
$\{\,\be_{\alpha}\,\}$. As $\be_{1}$ is presently a uniquely
defined vector, $\dot{{\bf e}}_{1}$ is fixed. This will thus have
an invariantly defined direction, say ${\bf X}$. For any given
tetrad choice, the components of this fixed direction in the
spatial $\be_{2}$-- and $\be_{3}$--directions are ${\bf X} \cdot
\be_{2}$ and ${\bf X} \cdot \be_{3}$, respectively. We have the
freedom to set one of these components (or any other quantity which
does not behave like a scalar under a rotation) to zero in ${\cal
{U}}$ by rotating the spatial triad $\{\,\be_{\alpha}\,\}$ in the
${\bf e}_{2}\,/\,\be_{3}$--plane. Alternatively, as can be seen
from the transformation property of $\Omega_1$,\footnote{See
\ct{bi:vE97}.}  we may choose to set $\Omega_{1} = 0$ in ${\cal
{U}}$ which fixes the $\be_{0}$--gradient of the rotation angle
$\varphi$; and then for example we choose either $n_{22}$ or
$n_{33}$ in the 3-space by fixing the $\be_{1}$--gradient of
$\varphi$.

For ease of notation we will use
\[
\Theta_\alpha := \sfrac{1}{3}\,\Th + \sigma_{\alpha\alpha }
~~~ (\mbox{no summation}) \ .
\]

\subsection{General perfect fluid}
This is the most general case to be considered here where we will
take throughout that $\dot{{\bf u}} \neq 0$ and
${\mbox{\boldmath{$\omega$}}} \neq 0$. For
${\mbox{\boldmath{$\omega$}}} = 0$ see section
\r{se:irrotfluid}, for $\dot{{\bf u}} = 0$ refer to section
\r{se:rotdust}. We assume a perfect fluid, but leave the {\em
equation of state\/} unspecified.

\subsubsection{Constraint analysis}
We proceed by fully fixing the tetrad freedom by choosing ${\bf
e}_{3}$ orthogonal to the projection of the fixed vector ${\bf X}$
in the $\be_{2}\,/\,\be_{3}$--plane. Hence, we rotate the spatial
triad such that
\beq
\Omega_{2} = \be_{3}\cdot\dot{\be}_{1} =
-\,\be_{1}\cdot\dot{\be}_{3} = 0 \ .
\eeq

Under the assumptions of PLRS symmetry (\r{eq:PLRS}), which imply
(\r{eq:zerotrn}), and with the above tetrad choice, the
$\dot{\bE}$--equations (\r{eq:01223}) and (\r{eq:02312}) yield the
new constraints
\bea
\l{eq:00}
\be_{2}(H_{11}) & = & 0 \\
\l{eq:0}
(a_{2}-n_{31})\,H_{11} & = & 0 \ ,
\eea
while from the $\dot{\bH}$--equations (\r{eq:01201}) and
(\r{eq:02303}), the new constraints
\bea
\l{eq:1}
\be_{2}(\mu) & = & 3\,(a_{2}-n_{31})\,E_{11} \\
\be_{2}(E_{11}) & = & \sfrac{1}{3}\,\be_{2}(\mu)
\eea
arise. The algebraic condition (\r{eq:0}) suggests that we
distinguish between two subcases according to
\[
\mbox{{\bf A}]} ~~ (a_{2}-n_{31}) = 0
~~~ \mbox{or} ~~~
\mbox{{\bf B}]} ~~ H_{11} = 0 \ .
\]

\paragraph*{A] $(a_{2}-n_{31}) = 0$:}

So now $(a_{2}+n_{31}) = 2\,a_{2}$. From the $\dot{\bw}$--equation
(\r{eq:Jac0012}) we see that $\be_{2}(\ud1) = 0$ and from
(\r{eq:1}) we have $\be_{2}(\mu) = 0$. The
$\dot{\mbox{\boldmath{$\sigma$}}}$--equation (\r{eq:R0102}) then
shows that also $\Omega_{3}\,\sigma_{11} = 0$ must hold,
providing a split into further subcases according to
\[
\mbox{{\bf A1}]} ~~ \Omega_{3} = 0
~~~ \mbox{or} ~~~
\mbox{{\bf A2}]} ~~ \sigma_{11} = 0 \ .
\]

\paragraph*{A1] $\Omega_{3} = 0$:}

This has the implications from (\r{eq:R0103}) that ${\bf
e}_{3}(\dot{u}_{1}) = 0 $, which implies from (\r{eq:Jac0013})
that $(a_{3}+n_{12}) = 0$, as we assumed $\dot{u}_{1} \neq
0$. We note that the following $\be_{2}$-- and ${\bf
e}_{3}$--gradients of certain quantities must vanish: from
(\r{eq:00}) and (\r{eq:02331}) we get that $\be_{2}(H_{11}) =
\be_{3} (H_{11}) = 0$ and from (\r{eq:01201}) -- (\r{eq:02302})
we find that $\be_{2}(E_{11}) = {\bf e}_{3}(E_{11}) = 0$ and $
\be_{2} (\mu) = \be_{3} (\mu) = 0$. Then (\r{eq:Jac1012}) and
(\r{eq:Jac1013}) reduce to
\beq
\l{eq:26}
\be_{2}(\Theta_{1}) = \be_{3}(\Theta_{1}) = 0 \ .
\eeq
Now we check the propagation property of the vanishing ${\bf
e}_{2}$-- and $\be_{3}$--gradients of the energy density $\mu$. We
use the commutator relations (\r{eq:commutat02}) and
(\r{eq:commutat03}) operating on $\mu$ and the energy conservation
equation (\r{eq:0b;b}) to show that ${\bf e}_{2}(\mu) =
\be_{3}(\mu) = 0$ if $(\mu+p)\,\be_{2}(\Th) = (\mu+p)\,\be_{3}(\Th)
= 0 $, since the momentum conservation equations (\r{eq:2b;b}) and
(\r{eq:3b;b}) must hold. Thus, if we want to stick to purely
cosmological solutions as we defined them in the introduction, we
must have ${\bf e}_{2}(\Th) =
\be_{3}(\Th) = 0$. Now this means that ${\bf e}_{2}(\omega_{1})
= \be_{3}(\omega_{1}) = 0$, which we obtain from the Ricci
identities (\r{eq:R0323}) and (\r{eq:R0223}), suitably combined
with (\r{eq:26}). This result is crucial because now we can show
that the solutions contained in the present PLRS subclass are {\em
not\/} cosmological ones. To do so we find the commutator
(\r{eq:commutat23}) most useful. We first apply this relation to
$\omega_{1}$, yielding
\beq
\l{eq:31}
2\,\Theta_{2}\,\omega_{1} - a_{1}\,n_{11} = 0\ ,
\eeq
where we have used the $\dot{\bw}$--equation (\r{eq:Jac0023}), and
the constraint on ${\bf e}_{1}(\omega_{1})$ given by
(\r{eq:Jac0123}). We then apply the commutator (\r{eq:commutat23})
to the energy density $\mu$, using (\r{eq:0b;b}), and find that
\beq
\l{eq:312}
2\,\Th\,\omega_{1}\,(\mu+p) - n_{11}\,\be_{1}(\mu) = 0 \ .
\eeq
Finally we apply the commutator (\r{eq:commutat23}) to the
electric Weyl curvature component $E_{11}$ and substitute from
(\r{eq:02323}), (\r{eq:12323}) and (\r{eq:312}) to get
\beq
\l{eq:30}
2\,\Th_{2}\,\omega_{1}\,[\,(\mu+p) - 3\,E_{11}\,]
+ 3\,a_{1}\,n_{11}\,E_{11} = 0 \ .
\eeq
We now use (\r{eq:31}) in (\r{eq:30}) to get
$\Theta_{2}\,\omega_{1}\,(\mu+p) = 0 \Rightarrow \Theta_{2} =
0$, as we assumed $(\mu+p) > 0$ and $\omega_{1} \neq
0$. Substituting back into (\r{eq:31}) now gives $a_{1}\,n_{11}
= 0$. We argue that this necessarily means $n_{11} = 0$. If
instead we started from $a_{1} = 0$, then (\r{eq:R0331}) shows
that $n_{11}\,\omega_{1} = 0 \Rightarrow n_{11} = 0$; hence,
$n_{11} = 0$ in any case. But then (\r{eq:Jac1123}) shows
$\Theta_{1} = 0$, and thus, since already $\Theta_{2} = 0$,
we find with the Raychaudhuri equation (\ref{eq:Raych}) that in
${\cal U}$
\beq
\Th = 0 \ ,
\eeq
violating the premise that our cosmology be an expanding one. We
conclude that there exist no cosmologically viable solutions in the
present PLRS subclass.

\paragraph*{A2] $\sigma_{11} = 0$:}

{}From the $\dot{\bs}$-- and $\dot{\bw}$--equations (\r{eq:R0103})
and (\r{eq:Jac0013}) we must have
\beq
\l{eq:3}
(a_{3}+n_{12})\,\ud1 - \Omega_{3}\,\omega_{1} = 0 \ ,
\eeq
and from the second Bianchi identities (\r{eq:03123}) and
(\r{eq:02331}),
\beq
\l{eq:4}
(a_{3}+n_{12})\,H_{11} + \Omega_{3}\,E_{11} = 0 \ .
\eeq
Combining these two equations, we get that either $(a_{3}+n_{12}) =
\Omega_{3} = 0$ which would be a subcase of {\bf A1]} and is
thus non-cosmological, or
\beq
\l{eq:23}
\ud1\,E_{11} + \omega_{1}\,H_{11} = 0 \ .
\eeq
An important algebraic relation is given by the ${\bH}$--constraint
(\r{eq:R0123}); that is
\beq
2\,(\ud1+a_{1})\,\omega_{1} + H_{11} = 0 \ .  \l{eq:11}
\eeq
The tetrad choice employed has the effect of generally eliminating
the gradients in the $\be_{2}$--direction of important scalars.  In
particular, from (\r{eq:R0131}) and (\r{eq:R0223}) we get
\beq
\l{eq:8}
\be_{2}(\omega_{1}) = 0 = \be_{3}(\Th) \ .
\eeq
Now, from  (\r{eq:Jac0012}) we get that
\beq
\l{eq:9}
\be_{2}(\ud1) = 0 \ ,
\eeq
and from (\r{eq:12302})
\beq
\l{eq:10}
\be_{2}(H_{11}) = 0 \ .
\eeq
{}From (\r{eq:01201}), suitably combined with (\r{eq:02303}), we
get
\beq
\l{eq:20}
\be_{2}(E_{11}) = \be_{2}(\mu) = 0 \ .
\eeq
If we now take the $\be_{2}$--gradient of equation (\r{eq:11}) and
substitute from the equations (\r{eq:8}), (\r{eq:9}) and (\r
{eq:10}), we get the useful result
\beq
\l{eq:12}
\be_{2}(a_{1}) = 0  \ .
\eeq
A relation for $\be_{3}(\omega_{1})$ is provided by (\r{eq:R0112})
and (\r{eq:R0323}):
\beq
\l{eq:14}
\be_{3}(\omega_{1}) + (a_{3}+n_{12})\,\omega_{1} = 0 \ .
\eeq
We get relations involving the $\be_{3}$--gradients of both
$E_{11}$ and $\mu$ from combining (\r{eq:03101}) and (\r{eq:02302})
suitably, thus yielding
\bea
\l{eq:17}
& & \be_{3}(E_{11}) - \sfrac13\,\be_{3}(\mu)
- 3\,\Omega_{3}\,H_{11} = 0 \ , \\
\l{eq:18}
& & \be_{3}(\mu) - 3\,(a_{3}+n_{12})\,E_{11}
+ 3\,\Omega_{3}\,H_{11} = 0 \ .
\eea
The commutators provide vital information here. We find from
(\r{eq:commutat23}) acting on $\omega_{1}$ that
\beq
\l{eq:19}
-\,\sfrac23\,\Th\,\omega_{1} + a_{1}\,n_{11}
= 2\,a_{2}\,(a_{3}+n_{12}) \ ,
\eeq
where we have used the $\dot{\bw}$--equation (\r{eq:Jac0023}), the
constraints on the gradients of $\omega_{1}$ provided by
(\r{eq:Jac0123}), (\r{eq:8}) and (\r{eq:14}), and then substituted
from (\r{eq:Jac1123}) with (\r{eq:R3112}) appropriately. We find
from the commutator (\r{eq:commutat23}) acting on the magnetic Weyl
curvature component $H_{11}$ that
\[
3\,H_{11}\,(-\,\sfrac23\,\Th\,\omega_{1}
+ a_{1}\,n_{11}) - n_{11}\,(\mu+p)\,\omega_{1}
- 12\,a_{2}\,E_{11}\,\Omega_{3} = 0 \ ,
\]
using a relation for $\be_{3}(H_{11})$ provided by (\r{eq:02331}),
noting the constraint (\r{eq:10}) on $\be_{2}(H_{11})$, the
constraint on $\be_{1}(H_{11})$ given by ({\r{eq:12301}), and then
using the $\dot{\bH}$--equation (\r{eq:02301}). We also needed
the first part of (\r{eq:20}) and (\r{eq:fomc2}). Combining the
above with (\r{eq:19}) we get the useful result
\beq
\l{eq:22}
n_{11}\,(\mu+p)\,\omega_{1} = -\,18\,a_{2}\,\Omega_{3}\,E_{11} \ ,
\eeq
where we have also used (\r{eq:4}). We now take the commutator
(\r{eq:commutat12}) operating on $\omega_{1}$ and substitute from
(\r{eq:8}), (\r{eq:9}), (\r{eq:12}), the vorticity constraint
equation (\r{eq:Jac0123}), and (\r{eq:14}) into the resultant
expression and we find that $n_{33}\,(a_{3}+n_{12})\,\omega_{1} =
0$. Now if $(a_{3}+n_{12})= 0$, then from (\r{eq:3}) we must have
$\Omega_{3} = 0$, and thus this would be a class already dealt with
in section {\bf A2]}; those were non-cosmological. So we conclude
that $n_{33} = 0$. We now take the $\be_{3}$--gradient of
(\r{eq:23}) and get the key result
\beq
\l{eq:25}
\ud1\,\be_{3}(\mu) = 0 ~~~ \Rightarrow ~~~
\be_{3}(\mu) = 0
\eeq
by using, in addition to (\r{eq:23}), equations (\r{eq:4}),
(\r{eq:14}) -- (\r{eq:18}), (\r{eq:Jac0013}) and (\r{eq:02331}). We
proceed to check the consistency of (\r{eq:17}) and (\r{eq:18}) in
the combined form $\be_{3}(E_{11}) - 3\,(a_{3}+n_{12})\,E_{11} =
0$. We propagate this along ${\bf u}$ and find
\beq
\l{eq:24}
\sfrac{3}{2}\,n_{11}\,[\,(a_{3}+n_{12})\,H_{11}
- \Omega_{3}\,E_{11}\,] = 0 \ ,
\eeq
where we have used the $\dot{\bE}$--equation (\r{eq:02323}),the
constraints on $\be_{2}(E_{11})$ and $\be_{2}(\mu)$ given
respectively by (\r{eq:20}) and (\r{eq:8}), the constraints on
$\be_{3}(H_{11})$ given by (\r{eq:02331}), and the Jacobi identity
(\r{eq:Jac1013}) together with (\r{eq:R0131}). The evolution along
${\bf u}$ of the $\be_{3}$--gradient of $E_{11}$ is obtained by
applying the commutator (\r{eq:commutat03}) to $E_{11}$ and using
the equations (\r{eq:02323}) with (\r{eq:R2331}). Now we may
rewrite (\r{eq:24}) by using (\r{eq:4}), obtaining
$n_{11}\,(a_{3}+n_{12})\,H_{11} = 0$. We show that this means that
$n_{11} = 0$. If not, then $(a_{3}+n_{12})\,H_{11} = 0$ which means
that $\Omega_{3}\,E_{11} = 0$ (from (\r{eq:4})), which then
contradicts (\r{eq:22}).

We have seen here that $n_{11} = 0$ is required. But now, again from
(\r{eq:22}), we must have $a_{2}\,\Omega_{3}\,E_{11} = 0$. So
either $a_{2}$ or $\Omega_{3}\,E_{11}$ vanishes. We consider these
possibilities below.

\begin{itemize}
\item  If $\Omega_{3}\,E_{11} = 0$, then from (\r{eq:4}) we
require $(a_{3}+n_{12})\,H_{11} = 0$. If now $H_{11} = 0$, then
(\r{eq:03112}) tells us that these solutions are not cosmological
since they require in ${\cal U}$ $(\mu+p)\,\omega_{1} = 0
\Rightarrow (\mu+p) = 0$. And if $(a_{3}+n_{12}) = 0$, then
from (\r{eq:19}) we get in ${\cal U}$ $\Th\,\omega_{1} = 0
\Rightarrow \Th = 0$, and we are clearly in the non-cosmological
realm again.

\item  If, on the other hand, $a_{2} = 0$, then again from
(\r{eq:19}) we must have in ${\cal U}$ $\Th\,\omega_{1} = 0
\Rightarrow \Th = 0$. So none of the solutions here are of
relevance to us.
\end{itemize}

\paragraph*{B] $H_{11} = 0$:}

Immediately we see from (\r{eq:03112}) that $(\mu+p) + 3\,E_{11}
= 0$.  Also, from (\r{eq:02301}), we get that $n_{11}\,E_{11}
= 0$. Now if $E_{11} = 0$, then $(\mu+p) = 0$ which is not
allowed. So $n_{11} = 0$. We also get from (\r{eq:03123}) that
$\Omega_{3}\,E_{11} = 0$, and once again we deduce that since
$E_{11} = 0 \Rightarrow (\mu+p) = 0$, it follows that we must
have $\Omega_{3} = 0$. Moreover $\be_{3}(\ud1) = 0 $ from
(\r{eq:R0103}); and $\be_{2}(\ud1) = 0 $ from (\r{eq:R0102}).
This tells us from (\r{eq:Jac0012}) that $(a_{2}-n_{31}) = 0$
and from (\r{eq:Jac0013}) that $(a_{3}+n_{12}) = 0$. And now
from (\r{eq:R3112}) we get that $\Th_{1}\,\omega_{1} = 0
\Rightarrow \Theta_{1} = 0$. We get from (\r{eq:01201}) and
(\r{eq:02303}) that $\be_{2}(E_{11}) = \be_{2}(\mu) = 0 $ and
from (\r{eq:03101}) and (\r{eq:02302}) that $\be_{3}(E_{11}) =
\be_{3}(\mu) = 0$. If we now take the above two relations for
the $\be_{2}$-- and $\be_{3}$--gradients of $\mu$ and substitute
them into the commutator (\r{eq:commutat23}) applied to $\mu$,
using (\r{eq:0b;b}), we get $\Th\,\omega_{1}\,(\mu+p) = 0$, and so
in ${\cal U}$
\beq
\Th\,(\mu+p) = 0 \ ;
\eeq
in other words, there are no solutions in this PLRS subclass that
are of cosmological interest.

\subsubsection{Summary}
We conclude that there exist no rotating and accelerating perfect
fluid cosmologies which are PLRS according to Definition \r{def:1}.

\subsection{Irrotational accelerating perfect fluid}
\l{se:irrotfluid}
These models have ${\mbox{\boldmath{$\omega$}}} = 0$. We assume
that $\dot{{\bf u}} \neq 0$. An immediate implication for the
commutation functions $n_{\alpha \beta}$, in addition to
(\r{eq:zerotrn}), is that $n_{11} = 0$, from (\r{eq:Jac0023}). Now
from the $\bH$--constraint (\r{eq:R0123}) it follows that $H_{11} =
0$, reducing the $\bH$--equation (\r{eq:02301}) to a trivial
statement. A useful point of departure is provided by the
$\dot{\bE}$--equations (\r{eq:01223}) and (\r{eq:03123}).  That is,
$\Omega_{2}\,E_{11} = \Omega_{3}\,E_{11} = 0$. A brief argument
below will show that this means that in ${\cal U}$
\beq
\l{eq:32}
\Omega_{2} = \Omega_{3} = 0 \ .
\eeq
\begin{itemize}
\item  The argument goes as follows: if $E_{11} = 0$, then from
(\r{eq:02323}) we have $(\mu+p)\,\sigma_{11} = 0 \Rightarrow
\sigma_{11} = 0$. We find that the following gradients vanish:
$\be_{1}(\mu) = 0 = \be_{1}(\Th)$ from (\r{eq:12323}) and
(\r{eq:R0331}), respectively. Now from the commutator relation
(\r{eq:commutat01}) acting on $\mu$ we get $\Th\,(\mu+p) = 0$; that
is, this is a non-cosmological subcase. In this little argument, we
have employed the assumption that the matter fluid has a {\em
barotropic equation of state\/}, $p = p(\mu)$.
\end{itemize}
So we must have (\r{eq:32}) holding, which then implies from
(\r{eq:R0102}) and (\r{eq:R0103}) that
\beq
\l{eq:33}
\be_{2}(\ud1) = \be_{3}(\ud1) = 0 \ .
\eeq
In turn the effect of the above is that $(a_{2}-n_{31}) =
(a_{3}+n_{12}) = 0 $ which derives from (\r{eq:Jac0012}) and
(\r{eq:Jac0013}). So we may write $(a_{2}+n_{31}) = 2\,a_{2}$ and
$(a_{3}-n_{12}) = 2\,a_{3}$.

\subsubsection{Tetrad choice and constraint analysis}
{}From the above we see that we are free to employ along ${\bf u}$
a {\em Fermi-propagated\/}
${\mbox{\boldmath{$\sigma$}}}$--eigentriad $\{\,\be_{\alpha}\,\}$
by setting $\Omega_{1} = \be_{0}(\varphi)$ via a spatial rotation
by an angle $\varphi$ in the ${\bf e}_{2}\,/\,\be_{3}$--plane so
that $\Omega_{1}^{\prime} = 0$. We note the following consistency
conditions:
\bea
n_{11} & = & 0 ~~~\mbox{(from (\r{eq:Jac1023}))}  \nonumber \\
(a_{2}-n_{31}) & = & 0 ~~~\mbox{(from (\r{eq:Jac1012}) and
(\r{eq:R0112}))}  \nonumber \\
(a_{3}+n_{12}) & = & 0 ~~~\mbox{(from (\r{eq:Jac1013}) and
(\r{eq:R0131}))} \ . \nonumber
\eea
We can further use the tetrad freedom on a 3-surface $x^{0} =
c^{0}$ to set $n_{33} = 0$. This condition is then preserved as we
can see from (\r{eq:Jac3012}). We can also set $a_{3} = 0$ on a
2-surface $x^{0} = c^{0}$, $x^{1} = c^{1}$. This we can do with
impunity since firstly $\be_{0}(a_{3}) = -\,\Theta_{2}\,a_{3}$ from
(\r{eq:Jac2023}) and (\r{eq:R0223}). Secondly $\be_{1}(a_{3}) =
a_{1}\,a_{3} $ from (\r{eq:Jac2123}) and (\r{eq:R2312}). We
conclude $a_{3} = 0$ everywhere. We may summarise: the only
non-zero commutation functions are $\ud1$, $\Theta_{1}$,
$\Theta_{2}$, $a_{1}$ and $a_{2}$. Of these quantities, only
$a_{2}$ does not have its $\be_{2}$-- and $\be_{3}$--gradients
vanishing; as can be seen from (\r{eq:33}), (\r{eq:R0112}) and
(\r{eq:R0323}), (\r{eq:R0131}) and (\r{eq:R0223}), (\r{eq:R2312})
and (\r{eq:R2331}). We may now proceed to use the remaining tetrad
freedom to set $\be_{3}(a_{2}) = 0$ on the line $x^{0} = c^{0}$,
$x^{1} = c^{1}$, $x^{2} = c^{2}$. This we can do by observing that
the $\be_{0}$--, $\be_{1}$-- and $\be_{2}$--derivatives of this
quantity are conserved respectively by applying the commutators
(\r{eq:commutat03}), (\r{eq:commutat31}) and (\r{eq:commutat23}) to
$a_{2}$. We also need (\r{eq:Jac3023}) and (\r{eq:R0323}), and
(\ref{eq:Jac3123}) and (\ref{eq:R2323}) to see this. Finally we set
$\be_{2}(a_{2}) = 0$ at an event $x^{i} = c^{i}$. This can be done
because firstly, the $\be_{0}$--derivative of $\be_{2}(a_{2})$ is
driven by a multiple of $\be_{2}(a_{2})$ (from (\r{eq:commutat02})
operating on $a_{2}$ and (\r{eq:Jac3023})); secondly, the
$\be_{1}$--derivative of $\be_{2}(a_{2})$ is driven by a multiple
of $\be_{2}(a_{2})$ (from (\r{eq:commutat12}) operating on $a_{2}$
and (\r{eq:Jac3123})); thirdly, the $\be_{2}$--derivative of
$\be_{2}(a_{2})$ is driven by a multiple of $\be_{2}(a_{2})$ (from
taking $\be_{2}$ of (\r{eq:R2323})); lastly, the
$\be_{3}$--derivative of $\be_{2}(a_{2})$ vanishes (from
(\r{eq:commutat23}) operating on $a_{2}$). The remaining
commutation functions have their gradients in the
$\be_{2}\,/\,\be_{3}$--plane vanishing, and the only equations
remaining constrain quantities in the invariant spatial
$\be_{1}$--direction (from (\r{eq:R0331}), (\r{eq:R3131}) and
(\r{eq:Jac3123})). The now algebraic relation (\r{eq:R2323})
determines $E_{11}$ in terms of $\mu$, say. The curvature variables
which remain non-zero in ${\cal U}$ are $p$, $\mu$, and
$E_{11}$. We find that all of these variables have their gradients
vanishing in the $\be_{2}\,/\,\be_{3}$--plane. This is obvious from
(\r{eq:2b;b}) and (\r{eq:3b;b}), and (\r{eq:01201}) combined with
(\r{eq:02303}), as well as (\r{eq:03101}) combined with
(\r{eq:02302}). The remaining constraints on the various
$\be_{1}$--gradients are given by (\r{eq:1b;b}) and
(\r{eq:12323}). The consistent solutions in ${\cal U}$ we so obtain
are the LRS class II solutions of Stewart and Ellis \ct{bi:SE68}.

\subsubsection{Summary}
Having assumed in this subsection a barotropic equation of state,
$p = p(\mu)$, we conclude that the only irrotational accelerating
perfect fluid cosmologies that are PLRS according to Definition
\r{def:1} are the expanding solutions in LRS class II of Stewart
and Ellis \ct{bi:SE68} (see also \ct{bi:vEE96} and \ct{bi:MB99}).

\subsection{Rotating dust}
\l{se:rotdust}
These models have $p = 0$. From part of the twice-contracted
Bianchi identities (momentum conservation equations),
$\be_{\alpha}(p) = 0 \Rightarrow \dot{u}^{\alpha} = 0$. We here
assume that ${\mbox{\boldmath{$\omega$}}} \neq 0$. Since we are
dealing with dust, we note that now our tetrad choice is that of
Ellis in \ct{bi:E67}.\footnote{Described in his Theorem 3.1.}
However, because we demand $\Th > 0$, we will refrain from
identifying his rotating dust solutions of LRS class I which
require $\Th = 0$. We immediately note from the Jacobi identities
(\r{eq:Jac0012}) and (\r{eq:Jac0013}) that in ${\cal U}$
$$\Omega_{2} = \Omega_{3} = 0 \ . $$
%

\subsubsection{Tetrad choice and constraint analysis}
{}We are free to propagate the spatial triad $\{\,\be_{\alpha}\,\}$
along ${\bf u}$ as anti-rotating by choosing
$(\omega_{1}+\Omega_{1}) = \be_{0}(\varphi)$, i.e.,
$(\omega_{1}+\Omega_{1})^{\prime} = 0$. We can further use the
tetrad freedom in ${\cal U}$ to set $n_{33}^{\prime} = 0$ by
choosing $\be_{1}(\varphi)$ such that $\be_{1}(\varphi) =
-\,n_{33}$. This quantity is conserved, as we can see from
(\r{eq:Jac3012}). But now we may not as yet proceed with a further
tetrad specification as in \ct{bi:E67}, where $(a_{3}-n_{12}) = 0$
on a 2-surface $x^{0} = c^{0}$, $x^{1} = c^{1}$, because this would
constrain the geometry. In particular, it actually requires
$\be_{2}(\omega_{1}) = 0$ for $(a_{3}-n_{12}) = 0$ to hold, using
(\r{eq:Jac2023}) and (\r{eq:R0223}). So we proceed by leaving the
tetrad freedom unfixed for now and see what the implications are
from consistency checks. From setting $\mbox{\boldmath{$\sigma$}}$
to be rotationally symmetric we do not get any immediate
constraints. But from setting $\bE$ to be rotationally symmetric we
get new constraints. Specifically we get
\begin{eqnarray*}
(a_{2}-n_{31})\,H_{11} & = & 0
~~~ \mbox{(from  the $\dot{\bf E}$--equation (\r{eq:01223}) combined
with (\r{eq:12302}))} \\
(a_{3}+n_{12})\,H_{11} & = & 0
~~~ \mbox{(from  the $\dot{\bf E}$--equation (\r{eq:03123}) combined
with (\r{eq:12303}))} \ .
\end{eqnarray*}
So naturally here we have a split into
\[
\mbox{{\bf A}]}~~ H_{11} = 0
~~~ \mbox{or} ~~~
\mbox{{\bf B}]} ~~ (a_{2}-n_{31}) = (a_{3}+n_{12}) = 0 \ .
\]

\paragraph*{A] $H_{11} = 0$:}

For $H_{11}$ to vanish, we must have from the $\dot{{\bf
H}}$--equation (\r{eq:02301}) that $n_{11}\,E_{11} = 0$.

\begin{itemize}
\item  A brief argument now shows that we are not interested in
$E_{11} = 0$. It goes as follows: if $E_{11} = 0$,
then from (\r{eq:02323}) we have $\mu\,\sigma_{11} = 0
\Rightarrow \sigma_{11} = 0$. For consistency we now
require from the $\dot{\mbox{\boldmath{$\sigma$}}}$--equations
(\r{eq:R0101}) and (\r{eq:R0202}) that $\omega_{1} = 0$; that is,
this case is dealt with elsewhere. In fact this leads to FLRW
solutions as we have already noted.
\end{itemize}

Then we conclude that $E_{11} \neq 0$ must hold and hence $n_{11} =
0$. This has the immediate implication from the $\bH$--constraint
equation (\r{eq:R0123}) that $a_{1}= 0$. Now we get from the second
Bianchi identities the following new constraints
\bea
\l{eq:38}
3\,(a_{2}-n_{31})\,E_{11} - \be_{2}(\mu) & = & 0
~~~ \mbox{(from  (\r{eq:01201}) and (\r{eq:02303}))} \\
\l{eq:39}
3\,(a_{3}+n_{12})\,E_{11} - \be_{3}(\mu) & = & 0
~~~ \mbox{(from (\r{eq:03101}) and (\r{eq:02302}))} \ .
\eea
We propagate (\r{eq:38}) along ${\bf u}$ twice, using the necessary
evolution equations. We use (\r{eq:Jac1012}), (\r{eq:02323}), and
(\r{eq:commutat02}) operating on $\mu$ with (\r{eq:0b;b}), to get
\beq
\l{eq:40}
\omega_{1}\,(a_{3}+n_{12})\,(3\,E_{11}-\mu)
+ 2\,\mu\,\be_{3}(\omega_{1}) = 0 \ .
\eeq
We now need (\r{eq:Jac0023}), (\r{eq:Jac1013}), (\r{eq:02323}),
(\r{eq:0b;b}), and (\r{eq:commutat03}) operating on $\omega_{1}$,
to get for consistency of (\r{eq:40}) that
\beq
\l{eq:41}
-\,\omega_{1}\,(a_{2}-n_{31})\,(3\,E_{11}-\mu)
+ 4\,\mu\,\be_{2}(\omega_{1}) = 0 \ .
\eeq
We propagate (\r{eq:39}) along ${\bf u}$ twice, using the necessary
evolution equations. These are (\r{eq:Jac1013}), (\r{eq:02323}),
and (\r{eq:commutat03}) operating on $\mu$ with (\r{eq:0b;b}),
which thus yields
\beq
\l{eq:42}
\omega_{1}\,(a_{2}-n_{31})\,(3\,E_{11}-\mu)
+ 2\,\mu\,\be_{2}(\omega_{1}) = 0 \ .
\eeq
We now need (\r{eq:Jac0023}), (\r{eq:Jac1012}), (\r{eq:02323}),
(\r{eq:0b;b}), and (\r{eq:commutat02}) operating on $\omega_{1}$ to
get for consistency of (\r{eq:42}) that
\beq
\l{eq:43}
\omega_{1}\,(a_{3}+n_{12})\,(3\,E_{11}-\mu)
- 4\,\mu\,\be_{3}(\omega_{1}) = 0 \ .
\eeq
We form linear combinations of the above four constraints to
facilitate our task at this point.
\begin{eqnarray*}
\mu\,\be_{3}(\omega_{1}) & = & 0
~~~ \mbox{(from  (\r{eq:40}) $-$ (\r{eq:43}))}
~~~ \Rightarrow ~~~ \be_{3}(\omega_{1}) = 0 \\
\mu\,\be_{2}(\omega_{1}) & = & 0
~~~ \mbox{(from  (\r{eq:41}) $+$ (\r{eq:42}))}
~~~ \Rightarrow ~~~ \be_{2}(\omega_{1}) = 0 \\
(a_{3}+n_{12})\,(3\,E_{11}-\mu) & = & 0
~~~ \mbox{(from  $2 \times$ (\r{eq:40}) $+$ (\r{eq:43}))} \\
(a_{2}-n_{31})(3\,E_{11}-\mu) & = & 0
~~~ \mbox{(from  (\r{eq:41}) $-\,2\times$ (\r{eq:42}))} \ .
\end{eqnarray*}

\begin{itemize}
\item  A brief argument now shows that $3\,E_{11} - \mu = 0$ is not
applicable. It goes as follows: we note that $\be_{2}(\omega_{1}) =
\be_{3}(\omega_{1}) = 0$. So from the commutator
(\r{eq:commutat23}) acting on $\omega_{1}$ and incorporating the
$\dot{\mbox{\boldmath{$\omega$}}}$--equation (\r{eq:Jac0023}) into
this, we get that $\Theta_{2} = 0$. The consistency of this
requires from (\r{eq:R0202}) that $3\,E_{11} - \mu
+ 3\,\omega_{1}^{2} = 0$, and if $3\,E_{11} - \mu = 0$, it must
necessarily follow that $\omega_{1} = 0$, which we excluded.
\end{itemize}

So we must conclude that $(a_{2}-n_{31}) = (a_{3}+n_{12}) =
0$. This is severely restrictive. We get from the commutator
(\r{eq:commutat23}) acting on $\omega_{1}$ that $\Theta_{2} =
0$, using the $\dot{\mbox{\boldmath{$\omega$}}}$--equation
(\r{eq:Jac0023}). And now from (\r{eq:Jac1123}) we must also have
$\Theta_{1} = 0$. These last two results, in particular, imply
that in ${\cal U}$
\beq
\Th = 0 \ ,
\eeq
which means that spacetimes in this PLRS subclass are
non-cosmological.

\paragraph*{{\bf B}] $(a_{2}-n_{31}) = (a_{3}+n_{12}) = 0$:}

Firstly we note from (\r{eq:R0112}) and (\r{eq:R0131}) that
\beq
\l{eq:44}
\be_{2}(\Theta_{1}) = \be_{3}(\Theta_{1}) = 0 \ .
\eeq
The critical constraints are obtained from the second Bianchi
identities: (\r{eq:01201}) combined with (\r{eq:02303}), and
(\r{eq:03101}) combined with (\r{eq:02302}) once again. They read,
respectively,
\begin{eqnarray*}
\be_{2}(E_{11}) = \be_{2}(\mu) & = & 0 \\
\be_{3}(E_{11}) = \be_{3}(\mu) & = & 0 \ .
\end{eqnarray*}
We proceed to check preservation along ${\bf u}$ of $\be_{2}(\mu) =
\be_{3}(\mu) = 0$ by using evolution equations obtained from the
commutator relations (\r{eq:commutat02}) and (\r{eq:commutat03})
when acting on $\mu$. We also require from this the relations given
by (\r{eq:0b;b}), (\r{eq:R0112}), (\r{eq:R0131}), (\r{eq:R0323})
and (\r{eq:R0223}). We get that
\begin{eqnarray*}
\be_{0}(\be_{2}(\mu)) & = & (\sfrac12\,\sigma_{11}-\sfrac43\,\Th)\,
\be_{2}(\mu) - 2\,\be_{3}(\omega_{1})
~~~ \Rightarrow ~~~ \be_{3}(\omega_{1}) = 0 \\
\be_{0}(\be_{3}(\mu)) & = & (\sfrac12\,\sigma_{11}-\sfrac43\,\Th) \,
\be_{3}(\mu) + 2\,\be_{2}(\omega_{1})
~~~ \Rightarrow ~~~ \be_{2}(\omega_{1}) = 0 \ .
\end{eqnarray*}
Taking these results and putting them back into (\r{eq:R0112}),
(\r{eq:R0131}), (\r{eq:R0323}) and (\r{eq:R0223}), and recalling
(\r{eq:44}), we have $\be_{2}(\Theta_{1}) = \be_{3}(\Theta_{1}) =
\be_{2}(\Theta_{2}) = \be_{3}(\Theta_{2}) = 0$. We note now from
(\r{eq:12302}) and (\r{eq:12303}) that it is apparent that the
gradients of $H_{11}$ are also degenerate in this fashion:
$\be_{2}(H_{11}) = \be_{3}(H_{11}) = 0$. We check preservation
along ${\bf u}$ of $\be_{2}(E_{11}) = \be_{3}(E_{11}) = 0$ by using
(\r{eq:12323}) and the commutators (\r{eq:commutat02}) and
(\r{eq:commutat03}) when acting on $E_{11}$. The results are that
\begin{eqnarray*}
\be_{0}(\be_{2}(E_{11})) & = & -\,\sfrac32\,H_{11}\,\be_{2}(n_{11})
\\
\be_{0}(\be_{3}(E_{11})) & = & -\,\sfrac32\,H_{11}\,\be_{3}(n_{11})
\ .
\end{eqnarray*}
If now $H_{11} = 0$, we have a case already dealt with in {\bf
A]}. So we must therefore conclude that $\be_{2}(n_{11}) =
\be_{3}(n_{11}) = 0$. Now here again crucial algebraic
constraints are obtained from the commutator (\r{eq:commutat23}).
Acting on $\omega_{1}$ we get
\beq
\l{eq:45}
\Theta_{2}\,\omega_{1} = a_{1}\,n_{11}
\eeq
from the $\dot{\mbox{\boldmath{$\omega$}}}$--equation
(\r{eq:Jac0023}) and the $\be_{1}(\omega_{1})$--constraint
(\r{eq:Jac0123}). Acting on $n_{11}$ we get
\beq
\l{eq:46}
n_{11}\,(\sigma_{11}\,\omega_{1} + a_{1}\,n_{11}) = 0
\eeq
from the evolution equation for $n_{11}$ (\r{eq:Jac1023}) and the
$\be_{1}(n_{11})$--constraint (\r{eq:Jac1123}). We show that we do
not get any relevant solutions here. We start with equation
(\r{eq:46}).
\begin{itemize}

\item  We first show that $n_{11} = 0$ leads to trivial
solutions. If $n_{11} = 0$, then from (\r{eq:Jac1123}) we must have
$\Theta_{1} = 0$ and then from (\r{eq:45}) this means that
$\Theta_{2} = 0$ and so this cannot be cosmological because
consequently in ${\cal U}$
\beq
\Th = 0 \ .
\eeq

\item  If, on the other hand, $n_{11} \neq 0$, and instead we have
from (\r{eq:46}) that $\sigma_{11}\,\omega_{1} + a_{1}\,n_{11} =
0$, then we immediately get from (\r{eq:45}) that $\Th\,\omega_{1}
= 0$; that is, in ${\cal U}$
\beq
\Th = 0 \ .
\eeq
So these too are not cosmological solutions.
\end{itemize}
%

\subsubsection{Summary}
We conclude that there are no rotating dust cosmologies which are
PLRS according to Definition \r{def:1}.

\subsection{Irrotational dust}
\l{se:irrotdust}
We shall now consider irrotational dust cosmologies under the PLRS
restrictions of Definition \r{def:1}: that is to say, cosmologies
where ${\mbox{\boldmath{$\omega$}}} = 0$ and $p = 0
\Rightarrow \dot{{\bf u}} = 0$. We can show that the only consistent
irrotational dust solutions of the EFE which have
$\mbox{\boldmath{$\sigma$}}$ and {\em both\/} ${\bf E}$ and ${\bf
H}$ rotationally symmetric in the same plane are known
solutions.\footnote{We may, with relative, ease generalise this
result to the situation where only $\mbox{\boldmath{$\sigma$}}$ and
${\bf H}$ are rotationally symmetric.} For example, the case
$\mbox{\boldmath{$\sigma$}} = 0$ leads to the FLRW dust solutions
\ct{bi:E67}. Proceeding, we find from (\r{eq:R0103}) and
(\r{eq:R0102}) that in ${\cal U}$
$$\Omega_{2} = \Omega_{3} = 0 \ .$$

\subsubsection{Tetrad choice and constraint analysis}
We note that for the models we are interested in here, we again
choose the ${\mbox{\boldmath{$\sigma$}}}$--eigentriad
$\{\,\be_{\alpha}\,\}$ to be {\em Fermi-propagated\/} along ${\bf
u}$ by setting $\Omega_{1} = \be_{0}(\varphi)$ via a spatial
rotation by an angle $\varphi$ in the ${\bf
e}_{2}\,/\,\be_{3}$--plane so that $\Omega_{1}^{\prime} = 0$. We
may also choose $n_{33}^{\prime} = 0 \Leftrightarrow
\be_{1}(\varphi) = -\,n_{33}$, since now the evolution equation
(\r{eq:Jac3012}) for $n_{33}$ becomes involutive. We will show that
the present PLRS subclass recovers the well known Szekeres dust
solutions \ct{bi:S75}. From the $\dot{\bE}$--equation
(\r{eq:01223}) we find $\be_{2}(H_{11}) =
\sfrac32\,(a_{2}-n_{31})\,H_{11}$, while (\r{eq:02312}) gives
$\be_{2}(H_{11}) = 0$, leading to the more useful algebraic result
\beq
\l{eq:47}
(a_{2}-n_{31})\,H_{11} = 0 \ .
\eeq
Then from (\r{eq:03123}) we find $\be_{3}(H_{11}) +
\sfrac32\,(a_{3}+n_{12})\,H_{11} = 0$, while (\r{eq:02331})
gives $\be_{3}(H_{11}) = 0$, now yielding the algebraic result
\beq
\l{eq:48}
(a_{3}+n_{12})\,H_{11} = 0 \ .
\eeq
Now the commutator (\r{eq:commutat23}) is helpful at this point. If
we operate on $H_{11}$, we find that $n_{11}\,\be_{1}(H_{11}) = 0$,
which subdivides the class into
\[
\mbox{{\bf A]}} ~~ n_{11} = 0
~~~\mbox{or} ~~~
\mbox{{\bf B}]} ~~ \be_{1}(H_{11}) = 0 \ .
\]

\paragraph*{A] $n_{11} = 0$:}

It is easy to see from (\r{eq:Jac1023}) that presently $n_{11} = 0$
is conserved along ${\bf u}$. Hence, the ${\bf H}$--constraint
(\r{eq:R0123}) gives $H_{11} = 0$, which the $\dot{{\bf
H}}$--equation (\r{eq:02301}) preserves; thus ${\bf H} = 0$. Now we
look at (\r{eq:02303}), which reads
\beq
\be_{2}(E_{11}) = \sfrac13\,\be_{2}(\mu) \ ,
\eeq
and we use (\r{eq:02323}), (\r{eq:R0323}), (\r{eq:0b;b}), and the
commutator (\r{eq:commutat02}) acting on $E_{11}$ and $\mu$, to
check the time evolution property of this last relation:
\[
\be_{0}(\be_{2}(E_{11}) - \sfrac13\,\be_{2}(\mu))
= -\,4\,\Theta_{2}\,[\,\be_{2}(E_{11}) - \sfrac13\,\be_{2}(\mu)\,]
\ .
\]
This is solved by $\be_{2}(E_{11}) = \sfrac13\,\be_{2}(\mu)$, and
we may now use (\r{eq:12331}) with the above to get
$\be_{2}(E_{11}) = \sfrac32\,(a_{2}-n_{31})\,E_{11} -
\sfrac16\,\be_{2}(\mu)$, which, in this manner, shows the
consistency with (\r{eq:01201}). Now we look at (\r{eq:02302}),
which reads
\beq
\be_{3}(E_{11}) = \sfrac13\,\be_{3}(\mu) \ ,
\eeq
and we use (\r{eq:02323}), (\r{eq:R0223}), (\r{eq:0b;b}), and the
commutator (\r{eq:commutat03}) acting on $E_{11}$ and $\mu$, to
check the time evolution property of this last relation:
\[
\be_{0}(\be_{3}(E_{11}) - \sfrac13\,\be_{3}(\mu))
= -\,4\,\Theta_{2}\,[\,\be_{3}(E_{11}) - \sfrac13\,\be_{3}(\mu)\,]
\ .
\]
This is solved by $\be_{3}(E_{11}) = \sfrac13\,\be_{3}(\mu)$, and
we may now use (\r{eq:12312}) with the above to get
$\be_{3}(E_{11}) = \sfrac32\, (a_{3}+n_{12})\,E_{11} -
\sfrac16\,\be_{3}(\mu)$, which, in this manner, shows the
consistency with (\r{eq:03101}). We have from (\r{eq:R2331}) and
(\r{eq:R2312}) that along ${\bf u}$, and, by substitution into the
relevant commutators, that everywhere $\be_{2}(a_{1}) =
\be_{3}(a_{1}) = 0$. We may at this point use some of the remaining
freedom in the commutators to (on a 2-surface $x^{0} = c^{0}$,
$x^{1} = c^{1}$) set $\be_{3}(a_{2}+n_{31}) =
\be_{2}(a_{3}-n_{12})$. This can be done because firstly 
\[ 
\be_{0}(\be_{3}(a_{2}+n_{31}) - \be_{2}(a_{3}-n_{12}))
= -\,2\,\Theta_{2}\,[\,\be_{3}(a_{2}+n_{31})
- \be_{2}(a_{3}-n_{12})\,] \ ,
\]
using the commutator (\r{eq:commutat03}) on $(a_{2}+n_{31})$ and
the commutator (\r{eq:commutat02}) on $(a_{3}-n_{12})$. Secondly,
\[
\be_{1}(\be_{3}(a_{2}+n_{31}) - \be_{2}(a_{3}-n_{12}))
= 2\,a_{1}\,[\,\be_{3}(a_{2}+n_{31})
- \be_{2}(a_{3}-n_{12})\,] \ ,
\]
using the commutator (\r{eq:commutat31}) on $(a_{2}+n_{31})$, the
commutator (\r{eq:commutat12}) on $(a_{3}-n_{12})$, and
(\r{eq:Jac3123}), (\r{eq:Jac2123}) and (\r{eq:Jac1123}). Thus
\beq
\l{eq:53}
\be_{3}(a_{2}+n_{31}) = \be_{2}(a_{3}-n_{12}) \ .
\eeq
There is still tetrad freedom remaining, but it is not obvious how
one could best utilise it. The commutation functions which remain
non-zero and the curvature variables they are coupled to in ${\cal
U}$ are $\Theta_{1}$, $\Theta_{2}$, $a_{\alpha}$, $n_{31}$,
$n_{12}$, $\mu$, and $E_{11}$. This may be recognised as the
Szekeres class I of dust solutions \ct{bi:S75} (and its subcase,
Ellis' dust spacetimes of LRS class II \ct{bi:E67}).  The evolution
equations and remaining constraint equations of these quantities
are also given there.

\paragraph*{{\bf B}] $\be_{1}(H_{11}) = 0$:}

Immediately we see from the $\be_{1}(H_{11})$--constraint
(\r{eq:12301}) that $a_{1}\,H_{11} = 0$, which provides us with the
subdivision
\[
\mbox{\bf B1]} ~~ a_{1} = 0
~~~ \mbox{or} ~~~
\mbox{\bf B2]} ~~ H_{11} = 0 \ .
\]

\paragraph*{B1] $a_{1} = 0$:}

We still have (\r{eq:47}) and (\r{eq:48}) holding, i.e.,
$(a_{2}-n_{31})\, H_{11} = (a_{3}+n_{12})\,H_{11} = 0$. Now if
$H_{11} = 0$, then we are dealing with {\bf B2]}. So instead here
we must have $(a_{2}-n_{31}) = (a_{3}+n_{12}) = 0$. We may use the
tetrad freedom to set $(a_{3}-n_{12}) = 0$. From (\r{eq:Jac2023})
and (\r{eq:R0223}) we have $(a_{3}-n_{12}) = 0$ without any further
new constraints. And now, from (\r{eq:Jac2123}), we have
$(a_{3}-n_{12}) = 0$ without any further new constraints. So this
choice is allowed. Thus we may state that
\[
a_{3} = n_{12} = 0
~~~ \mbox{and} ~~~
(a_{2}+n_{31}) = 2\,a_{2} \ .
\]
Now from (\r{eq:01201}) combined with (\r{eq:02303}),
(\r{eq:03101}) combined with (\r{eq:02302}), the commutator
(\r{eq:commutat23}) operating on $\mu$, and then using
(\r{eq:01202}), we get that $\be_{\alpha}(\mu) =
\be_{\alpha}(E_{11}) = 0$. Also, from (\r{eq:R0112}) and
(\r{eq:R0323}) we have $\be_{2}(\Theta_{1}) = \be_{2}(\Theta_{2}) =
0 \Rightarrow \be_{2}(\Th) = \be_{2}(\sigma_{11}) = 0$, while from
(\r{eq:R0131}) and (\r{eq:R0223}) follows $\be_{3}(\Theta_{1}) =
{\bf e}_{3}(\Theta_{2}) = 0 \Rightarrow
\be_{3}(\Th) = \be_{3}(\sigma_{11}) = 0$,  which then, from the
commutator (\r{eq:commutat23}) acting on $\Th$ and $\sigma_{11}$,
gives $\be_{\alpha}(\Th) = {\bf e}_{\alpha}(\sigma_{11}) =
0$. Moreover, it is clear from (\r{eq:01223}) and (\r{eq:03123})
that $\be_{\alpha} (H_{11}) = 0$. The constraints
(\r{eq:Jac1123}), (\r{eq:R2312}) and (\r{eq:R2331}) imply
$\be_{\alpha}(n_{11}) = 0$. By applying the commutators
(\r{eq:commutat01}), (\r{eq:commutat02}) and (\r{eq:commutat03}) on
the variables $f \in \{\,\mu, \,E_{11}, \,H_{11}, \,\Th,
\,\sigma_{11}, \,n_{11}\,\}$, and utilising their respective
evolution equations (\r{eq:0b;b}), (\r{eq:02323}), (\r{eq:02301}),
(\r{eq:R0101}), (\r{eq:R0202}) and (\r{eq:Jac1023}), we can show
that $\be_{\alpha}(f) = 0$.  Now from (\r{eq:Jac3123}) we have
$\be_{1}(a_{2}) = 0$, which allows us to use the remaining freedom
to set $\be_{3}(a_{2}) = 0$.  We can do this because
\[
\be_{0}(\be_{3}(a_{2})) = -\,2\,\Theta_{2}\,\be_{3}(a_{2})
~~~ \mbox{(from  (\r{eq:commutat03}) on $a_{2}$,
 (\r{eq:Jac3023}) and  (\r{eq:R0223}))} \ ,
\]
\[
\be_{1}(\be_{3}(a_{2})) = 0
~~~ \mbox{(from  (\r{eq:commutat31}) on $a_{2}$ and
(\r{eq:Jac3123}))} \ ,
\]
and
\[
\be_{2}(\be_{3}(a_{2})) = 6\,a_{2}\,\be_{3}(a_{2})
~~~ \mbox{(from  (\r{eq:commutat23}) on $a_{2}$,
(\r{eq:R2323}) and (\r{eq:Jac3123}))} \ .
\]
Finally we set $\be_{2}(a_{2}) = 0$ at an event $x^{i} =
c^{i}$. This can be done because firstly, the $\be_{0}$--derivative
of $\be_{2}(a_{2})$ is given by a multiple of $\be_{2}(a_{2})$
(from (\r{eq:commutat02}) operating on $a_{2}$ and
(\r{eq:Jac3023})); secondly, the $\be_{1}$--derivative of
$\be_{2}(a_{2})$ vanishes (from (\r{eq:commutat12}) operating on
$a_{2}$ and (\r{eq:Jac3123})); thirdly, the $\be_{2}$--derivative
of $\be_{2}(a_{2})$ is given by a multiple of $\be_{2}(a_{2})$
(from taking $\be_{2}$ of (\r{eq:R2323})); lastly, the
$\be_{3}$--derivative of $\be_{2}(a_{2})$ vanishes (from
(\r{eq:commutat23}) operating on $a_{2}$).  The remaining non-zero
commutation functions and curvature variables they are coupled to
in ${\cal U}$ are $\mu$, $\Theta_{1}$, $\Theta_{2}$, $a_{2}$,
$n_{11}$, $E_{11}$ and $H_{11}$. The remaining constraints
(\r{eq:R0123}), (\r{eq:R3131}) and (\r{eq:R2323})) are all
algebraic relations. Thus the presently obtained consistent
solutions constitute the subclass of spatially homogeneous
cosmologies within the dust LRS class II of Ellis \ct{bi:E67}.

\paragraph*{B2] $H_{11} = 0$:}

{}From the $\dot{{\bf H}}$--equation (\r{eq:02301}) it follows that
$n_{11}\,\sigma_{11} = 0$.  Now if $n_{11} = 0$ then we are dealing
with case {\bf A]}, and if in ${\cal U}$ $\sigma_{11} = 0$, these
are the FLRW dust solutions and are well known.

\subsubsection{Summary}
We conclude that the only non-trivial irrotational dust cosmologies
that are PLRS according to Definition \r{def:1} are known. These
are the Szekeres dust cosmologies \ct{bi:S75}, which are strictly
PLRS, and Ellis' dust spacetimes in LRS class II \ct{bi:E67}.
Moreover, this last study demonstrates that local rotational
symmetry of an irrotational dust cosmology results if, in an open
neighbourhood ${\cal U}$ of an event of ${\cal M}$, all tensors
algebraically determined from the Riemann curvature tensor and all
its covariant derivatives up to only {\em second order\/} are
rotationally symmetric, generalising a corresponding result given
in \ct{bi:E67}. This property may be traced back to the fact that a
rotationally symmetric covariant derivative $\nabla_{1}\be_{1}$,
i.e.
\[
\Gamma_{211} = \be_{2} \cdot \nabla_{1}\be_{1} = 0
= \be_{3} \cdot\nabla_{1}\be_{1} = \Gamma_{311} \ ,
\]   
corresponds by (\r{eq:commutat31}) and (\r{eq:commutat12}) to 
\[
0 = (a_{2}-n_{31}) = (a_{3}+n_{12}) \ ,
\]
eliminating the Szekeres dust models from the above consistent set
and reducing to the LRS solutions. Note that these conditions also
apply when $p \neq 0 \Leftrightarrow \ud1 \neq 0$ (cf. subsection
\r{se:irrotfluid}).

\section{Conclusion}
The only consistent solutions for the general class of perfect
fluid cosmologies with all tensors rotationally symmetric about the
same spatial axis are known solutions. All of these solutions are
either LRS or they belong to the Szekeres class of dust
cosmologies. Thus our work may be viewed as a form of
classification scheme for inhomogenous cosmologies which
incorporates the Szekeres dust solutions. This may be contrasted
with the scheme developed by Szafron and Collins, where
restrictions placed on submanifolds achieve this result
\cite{bi:CS79a, bi:SC79, bi:CS79b}. A main result of the present
work is that for all {\em spatially inhomogeneous\/} perfect fluid
cosmologies that are PLRS according to Definition \r{def:1} we have
in ${\cal U}$
\[
{\bf H} = 0 \ .
\]
This generalises similar results obtained for the Szekeres dust
cosmologies in \ct{bi:GW82}, and for perfect fluid cosmologies in
LRS class II in \ct{bi:vEE96}. It has also been demonstrated that
perfect fluid cosmologies are LRS if tensors and their covariant
derivatives up to {\em second order\/} are rotationally symmetric,
generalising a similar result obtained in \ct{bi:E67}.

There are no PLRS perfect fluid cosmologies which are
rotating. Also, the cases which have vanishing fluid shear are
fairly trivial: they are either not cosmologies at all in our
understanding or they are of the simple FLRW kind. This last result
requires the assumption of a barotropic equation of state $p =
p(\mu)$ in the irrotational accelerating perfect fluid case
(although it is probably possible to relax this requirement).

We have seen that the only strictly PLRS solutions are the Szekeres
dust cosmologies. In later work we intend to relax the requirements
of Definition \r{def:1} to see what partial symmetry features
result. One possibility is to keep $\mbox{\boldmath{$\sigma$}}$ and
$\bE$ rotationally symmetric about the same spatial axis, but leave
$\bH$ unrestricted such that a uniquely defined tetrad
$\{\,\be_{a}\,\}$ does exist. An example of a spatially
inhomogeneous perfect fluid cosmology with coinciding eigenvalues
of $\mbox{\boldmath{$\sigma$}}$ and non-zero $\bE$ and $\bH$ is
provided by the diagonal Abelian $G_{2}$ solutions with radiation
equation of state by Senovilla \ct{bi:S90}.

\section*{Acknowledgements}
HvE was supported by the Deutsche Forschungsgemeinschaft (DFG). MM
was supported by the Royal Swedish Academy of Sciences (KVA). We
would like to thank Roy Maartens for some good comments on an
earlier draft and our referees for their helpful
recommendations. In parts of this work the algebraic computing
package {\tt CLASSI} was employed.

\appendix
\section*{Appendix}
\section{Evolution and constraint equations}
\l{ap:fullset}
{\bf Restrictions imposed}: All $1+3$ ONT equations we list in the
following are written out for cosmologies with a perfect fluid
matter source that are subject to the specific restrictions that
\begin{itemize}
\item 
a fluid-comoving $1+3$ ONT $\{\,\be_{a}\,\} = \{\,{\bf u},
\,\be_{\alpha}\,\}$ is chosen which {\em diagonalises\/}
${\mbox{\boldmath{$\sigma$}}}$,

\item 
${\mbox{\boldmath{$\sigma$}}}$ is {\em rotationally symmetric\/} in
the $\be_{2}\,/\,\be_{3}$--plane; meaning, $\sigma_{22} =
\sigma_{33}\ \left(= -\,\sfrac{1}{2}\,\sigma_{11}\right)$,

\item 
$\omega_{2} = \omega_{3} = 0$, that is,
${\mbox{\boldmath{$\omega$}}} \parallel \be_{1}$,

\item 
$\dot{u}_{2} = \dot{u}_{3} = 0$; equivalently,
$\dot{{\bf u}} \parallel \be_{1}$.
\end{itemize}
For brevity we use the variables $\Th_{1}$ and $\Th_{2}$ in some
sets of equations while we use $\sigma_{11}$ and $\Th$ in other
areas where the degeneracy in the eigenvalues of
${\mbox{\boldmath{$\sigma$}}}$ allows for some simplification. The
relations
\[
\sigma_{11} = \sfrac23\,(\Th_{1} - \Th_{2})
~~~ \mbox{and} ~~~
\Th = \Th_{1} + 2\,\Th_{2}
\]
yield the transformation from the former to the latter set of
variables.

The $1+3$ ONT equations for perfect fluid cosmologies in
fluid-comoving description {\em without\/} the specialisation
stated can be accessed online at the URL \ct{bi:vE97}.

\subsection{Rotational tetrad freedom}
For coinciding eigenvalues of the fluid shear tensor
$\mbox{\boldmath{$\sigma$}}$ such that $\sigma_{22} = \sigma_{33}$,
its eigentetrad $\{\,\be_{a}\,\}$ is defined only up to spatial
rotations ${\mbox{\boldmath{$\Lambda$}}} =
{\mbox{\boldmath{$\Lambda$}}}(x^{i})$ in the
$\be_{2}\,/\,\be_{3}$--plane. These may be represented by
\beq
\l{eq:rotation}
\Lambda^{-1}{}_{a{\prime}}{}^{a}
= \left(
\begin{array}{rrrr}
1 & 0 & 0 & 0 \\
0 & 1 & 0 & 0 \\
0 & 0 & \cos \varphi  & \sin \varphi  \\
0 & 0 & -\sin \varphi  & \cos \varphi
\end{array}
\right) \ ,
\eeq
with, in general, $\be_{0}(\varphi) \neq 0 \neq \be_{1}(\varphi)$.
The members of the $\mbox{\boldmath{$\sigma$}}$--eigentetrad then
transform according to
\[
\be_{a{\prime}} \rightarrow \Lambda^{-1}{}_{a{\prime}}{}^{a}\,
\be_{a} \ .
\]
%

\subsection{Commutators}
In explicit form the commutators read
\bea
\l{eq:commutat01}
[\,\be_{0}, \,\be_{1}\,]\,(f)
& = & \dot{u}_{1}\,\be_{0}(f)
- \Th_{1}\,\be_{1}(f)
- \Omega_{3}\,\be_{2}(f) + \Omega_{2}\,\be_{3}(f) \\
\l{eq:commutat02}
[\,\be_{0}, \,\be_{2}\,]\,(f)
& = & \Omega_{3}\,\be_{1}(f)
- \Th_{2}\,\be_{2}(f)
- (\omega_{1}+\Omega_{1})\,\be_{3}(f) \\
\l{eq:commutat03}
[\,\be_{0}, \,\be_{3}\,]\,(f)
& = & -\,\Omega_{2}\,\be_{1}(f) + (\omega_{1}+\Omega_{1})\,
\be_{2}(f) - \Th_{2}\,\be_{3}(f) \\
\l{eq:commutat23}
[\,\be_{2}, \,\be_{3}\,]\,(f)
& = & 2\,\omega_{1}\,\be_{0}(f) + n_{11}\,\be_{1}(f)
- (a_{3}-n_{12})\,\be_{2}(f) + (a_{2}+n_{31})\,\be_{3}(f)
\\
\l{eq:commutat31}
[\,\be_{3}, \,\be_{1}\,]\,(f)
& = & (a_{3}+n_{12})\,\be_{1}(f) + n_{22}\,\be_{2}(f)
- (a_{1}-n_{23})\,\be_{3}(f) \\
\l{eq:commutat12}
[\,\be_{1}, \,\be_{2}\,]\,(f)
& =  & -\,(a_{2}-n_{31})\,\be_{1}(f) + (a_{1}+n_{23})\,
\be_{2}(f) + n_{33}\,\be_{3}(f) \ .
\eea
%

\subsection{Evolution equations}
\subsubsection{Energy density evolution}
Conservation of energy requires
\beq
\l{eq:0b;b}
\be_{0}(\mu) = -\,\Th\,(\mu+p) \ .
\eeq
%

\subsubsection{Fluid shear and expansion evolution in terms of
${\bf E}$}
%
\bea
\l{eq:R0101}
R_{0101} & = & -\,(\be_{0}+\Th_{1})\,
(\Th_{1})
+ (\be_{1}+\dot{u}_{1})\,(\dot{u}_{1})
\ = \ E_{11} + \sfrac{1}{6}\,(\mu+3p) \\
\l{eq:R0202}
R_{0202} & = & -\,(\be_{0}+\Th_{2})\,
(\Th_{2})
+ \omega_{1}^{2} - (a_{1}+n_{23})\,\dot{u}_{1}
\ = \ E_{22} + \sfrac{1}{6}\,(\mu+3p) \\
\l{eq:R0303}
R_{0303} & = & -\,(\be_{0}+\Th_{2})\,
(\Th_{2})
+ \omega_{1}^{2} - (a_{1}-n_{23})\,\dot{u}_{1}
\ = \ E_{33} + \sfrac{1}{6}\,(\mu+3p) \ .
\eea
%

\subsubsection{Fluid expansion evolution}
The expansion evolution is obtained by writing out
$R_{0}{}^{\alpha}{}_{0\alpha}$:
\beq
\l{eq:Raych}
\be_{0}(\Th) = -\,\sfrac{1}{3}\,\Th^{2}
+ (\be_{1}+\dot{u}_{1}-2\,{a}_{1})\,(\dot{u}_{1})
- \sfrac{3}{2}\,\sigma_{11}^{2} + 2\,\omega_{1}^{2}
- \sfrac{1}{2}\,(\mu+3p) \ .
\eeq
%

\subsubsection{Fluid vorticity and spatial commutation function
evolution}
%
\bea
\l{eq:Jac0023}
\left({}^{~\,0}_{0 2 3}\right) \hspace{14mm}
\be_0(\omega_{1}) & = & -\,2\,\Th_{2}\,
\omega_{1} - \sfrac{1}{2}\,n_{11}\,\dot{u}_{1} \\
\l{eq:Jac2012}
\left({}^{~\,2}_{0 1 2}\right) \hspace{5mm}
\be_{0}(a_{1}+n_{23})
& = & -\,\Th_{1}\,(a_{1}+n_{23})
- (\be_{1}+\ud1)\,(\Th_{2})
+ (\be_{2}-a_{2}+n_{31})\,(\Omega_{3})
\nonumber \\
& & \hspace*{1cm} + \ (\omega_{1}+\Omega_{1})\,(n_{22}-n_{33})
+ \Omega_{2}\,(a_{3}-n_{12}) \\
\l{eq:Jac3013}
\left({}^{~\,3}_{0 1 3}\right) \hspace{5mm}
\be_{0}(a_{1}-n_{23})
& = & -\,\Th_{1}\,(a_{1}-n_{23})
- (\be_{1}+\ud1)\,(\Th_{2})
- (\be_{3}-a_{3}-n_{12})\,(\Omega_{2})
\nonumber \\
& & \hspace*{1cm} -  \ (\omega_{1}+\Omega_{1})\,(n_{22}-n_{33})
-  \Omega_{3}\,(a_{2}+n_{31}) \\
\l{eq:Jac1012}
\left({}^{~\,1}_{0 1 2}\right) \hspace{5mm}
\be_{0}(a_{2}-n_{31})
& = & -\,\Th_{2}\,(a_{2}-n_{31})
- \be_{2}(\Th_{1})
- (\be_{1}+\ud1-a_{1}-n_{23})(\Omega_{3})
\nonumber \\
& & \hspace*{1cm} - \ (\omega_{1}+\Omega_{1})\,(a_{3}+n_{12})
- \Omega_{2}\,(n_{33}-n_{11}) \\
\l{eq:Jac1013}
\left({}^{~\,1}_{0 1 3}\right) \hspace{5mm}
\be_{0}(a_{3}+n_{12})
& = & -\,\Th_{2}\,(a_{3}+n_{12})
- \be_{3}(\Th_{1})
+ (\be_{1}+\ud1-a_{1}+n_{23})\,(\Omega_{2})
\nonumber \\
& & \hspace*{1cm} + \ (\omega_{1}+\Omega_{1})\,(a_{2}-n_{31})
+  \Omega_{3}\,(n_{11}-n_{22}) \\
\l{eq:Jac3023}
\left({}^{~\,3}_{0 2 3}\right) \hspace{5mm}
\be_{0}(a_{2}+n_{31})
& = & -\,(\be_{2}+a_{2}+n_{31})\,(\Th_{2})
+ (\be_{3}-a_{3}+n_{12})\,(\omega_{1}+\Omega_{1}) \nonumber \\
& & \hspace*{1cm} + \ \Omega_{3}\,(a_{1}-n_{23})
+  \Omega_{2}\,(n_{33}-n_{11}) \\
\l{eq:Jac2023}
\left({}^{~\,2}_{0 2 3}\right) \hspace{5mm}
\be_{0}(a_{3}-n_{12})
& = & - \,(\be_{3}+a_{3}-n_{12})\,(\Th_{2})
- (\be_{2}-a_{2}-n_{31})\,(\omega_{1}+\Omega_{1})
\nonumber \\
& & \hspace*{1cm} - \ \Omega_{2}\,(a_{1}+n_{23})
- \Omega_{3}\,(n_{11}-n_{22}) \\
\l{eq:Jac1023}
\left({}^{~\,1}_{0 2 3}\right) \hspace{13mm}
\be_{0}(n_{11}) & = & -\,(\sfrac{1}{3}\,\Th-2\,\sigma_{11})\,n_{11}
- (\be_{2}-2\,n_{31})\,(\Omega_{2})
- (\be_{3}+2\,n_{12})\,(\Omega_{3}) \\
\l{eq:Jac2013}
\left({}^{~\,2}_{0 1 3}\right) \hspace{13mm}
\be_{0}(n_{22}) & = & -\,\Th_{1}\,n_{22}
- (\be_{1}+\ud1+2\,n_{23})\,(\omega_{1}+\Omega_{1})
- (\be_{3}-2\,n_{12})\,(\Omega_{3}) \\
\l{eq:Jac3012}
\left({}^{~\,3}_{0 1 2}\right) \hspace{13mm}
\be_{0}(n_{33}) & = & -\,\Th_{1}\,n_{33}
- (\be_{1}+\ud1-2\,n_{23})\, (\omega_{1}+\Omega_{1})
- (\be_{2}+2\,n_{31})\,(\Omega_{2}) \ .
\eea
%

\subsubsection{Evolution equations for ${\bf E}$ and ${\bf H}$}
%
\bea
\framebox[1.4cm]{[023]\,23} && ~~~ (\be_{0} + 3 \Th_{2}) ({E}_{11})
= (\be_{2} - {a}_{2} - {n}_{31}) ({H}_{31}) - (\be%
_{3} - {a}_{3} + {n}_{12}) ({H}_{12})  \nonumber  \l{eq:02323} \\
&& \hspace*{1cm} - \sfrac12 (\mu + p) {\sigma}_{11} + 2 {\Omega}_{2} {E}_{31}
- 2 {\Omega}_{3} {E}_{12}  \nonumber \\
&& \hspace*{1cm} - \sfrac32 {n}_{11} {H}_{11} + \sfrac12 ({n}_{22}- {n}_{33}) (%
{H}_{22}- {H}_{33}) + 2 {n}_{23} {H}_{23} \\
\framebox[1.4cm]{[031]\,31} && ~~~ (\be_{0} + {\Th}) ({E}_{22}) = - (%
\be_{1} + 2 {\dot{u}}_{1} - {a}_{1} + {n}_{23}) ({H}_{23}) + (\be%
_{3} - {a}_{3} - {n}_{12}) ({H}_{12})  \nonumber  \l{eq:03131} \\
&& \hspace*{1cm} + \sfrac14 (\mu + p) {\sigma}_{11} + \sfrac32 {\sigma}_{11} {E%
}_{33} + 2 {\Omega}_{3} {E}_{12} - ({\omega}_{1} + 2 {\Omega}_{1}) {E}_{23}
\nonumber \\
&& \hspace*{1cm} - \sfrac32 {n}_{22} {H}_{22} + \sfrac12 ({n}_{33}- {n}_{11}) (%
{H}_{33} - {H}_{11}) + 2 {n}_{31} {H}_{31} \\
\framebox[1.4cm]{[012]\,12} && ~~~ (\be_{0} + {\Th}) ({E}_{33}) = (%
\be_{1} + 2 {\dot{u}}_{1} - {a}_{1} - {n}_{23}) ({H}_{23}) - (\be%
_{2} - {a}_{2} + {n}_{31}) ({H}_{31})  \nonumber  \l{eq:01212} \\
&& \hspace*{1cm} + \sfrac14 (\mu + p) {\sigma}_{11} + \sfrac32 {\sigma}_{11} {E%
}_{22} - 2 {\Omega}_{2} {E}_{31} + ( {\omega}_{1} + 2 {\Omega}_{1}) {E}_{23}
\nonumber \\
&& \hspace*{1cm} - \sfrac32 {n}_{33} {H}_{33} + \sfrac12 ({n}_{11}- {n}_{22}) (%
{H}_{11} - {H}_{22}) + 2 {n}_{12} {H}_{12} \\
\framebox[1.4cm]{[031]\,12} && ~~~ (\be_{0} + {\Th} + \sfrac32 {\sigma}%
_{11}) ({E}_{23}) = - (\be_{1} + {\dot{u}}_{1} - {a}_{1} + {n}_{23}) ({H}%
_{33}) + (\be_{3} - 2 {a}_{3} - 2 {n}_{12}) ({H}_{31})  \nonumber
\l{eq:03112} \\
&& \hspace*{1cm} + \sfrac12 (\mu + p) {\omega}_{1} - {\omega}_{1} ({E}_{33} -
{E}_{11}) - {\Omega}_{1} ({E}_{33} - {E}_{22}) - {\Omega}_{2} {E}_{12} + {%
\Omega}_{3} {E}_{31}  \nonumber \\
&& \hspace*{1cm} - \sfrac12 (3 {n}_{22} + {n}_{33} - {n}_{11}) {H}_{23} - ( {a%
}_{2} + {n}_{31}) {H}_{12} - ({a}_{1} - {n}_{23}) {H}_{11} + {\dot{u}}_{1} {H%
}_{22} \\
\framebox[1.4cm]{[012]\,31} && ~~~ (\be_{0} + {\Th} + \sfrac32 {\sigma}%
_{11}) ({E}_{23}) = (\be_{1} + {\dot{u}}_{1} - {a}_{1} - {n}_{23}) ({H}%
_{22}) - (\be_{2} - 2 {a}_{2} + 2 {n}_{31}) ({H}_{12})  \nonumber
\l{eq:01231} \\
&& \hspace*{1cm} -\sfrac12 (\mu + p) {\omega}_{1} + {\omega}_{1} ({E}_{22} - {%
E}_{11}) - {\Omega}_{1} ({E}_{33} - {E}_{22}) + {\Omega}_{3} {E}_{31} - {%
\Omega}_{2} {E}_{12}  \nonumber \\
&& \hspace*{1cm} - \sfrac12 (3 {n}_{33} - {n}_{11} + {n}_{22}) {H}_{23} + ({a}%
_{3} - {n}_{12}) {H}_{31} + ({a}_{1} + {n}_{23}) {H}_{11} - {\dot{u}}_{1} {H}%
_{33} \\
\framebox[1.4cm]{[012]\,23} && ~~~ (\be_{0} + {\Th}) ({E}_{31}) = (%
\be_{1} + {\dot{u}}_{1} - 2 {a}_{1} - 2 {n}_{23}) ({H}_{12}) - (\be%
_{2} - {a}_{2} + {n}_{31}) ({H}_{11})  \nonumber  \l{eq:01223} \\
&& \hspace*{1cm} - {\Omega}_{2} ({E}_{11} - {E}_{33}) - {\Omega}_{3} {E}%
_{23} + ( 2 {\omega}_{1} + {\Omega}_{1}) {E}_{12}  \nonumber \\
&& \hspace*{1cm} - \sfrac12 (3 {n}_{33} + {n}_{11} - {n}_{22}) {H}_{31} - ( {a%
}_{3} + {n}_{12}) {H}_{23} - ({a}_{2} - {n}_{31}) {H}_{22} \\
\framebox[1.4cm]{[023]\,12} && ~~~ (\be_{0} + 3 \Th_{2}) ({E}_{31})
= (\be_{2} - {a}_{2} - {n}_{31}) ({H}_{33}) - (\be%
_{3} - 2 {a}_{3} + 2 {n}_{12}) ({H}_{23})  \nonumber  \l{eq:02312} \\
&& \hspace*{1cm} - {\Omega}_{2} ({E}_{11} - {E}_{33}) + ( {\Omega}_{1} - {%
\omega}_{1}) {E}_{12} - {\Omega}_{3} {E}_{23}  \nonumber \\
&& \hspace*{1cm} - \sfrac12 (3 {n}_{11} - {n}_{22} + {n}_{33}) {H}_{31} + ({%
\dot{u}}_{1} + {a}_{1} - {n}_{23}) {H}_{12} + ({a}_{2} + {n}_{31}) {H}_{22}
\\
\framebox[1.4cm]{[031]\,23} && ~~~ (\be_{0} + {\Th} ) ({E}_{12}) = - (%
\be_{1} + {\dot{u}}_{1} - 2 {a}_{1} + 2 {n}_{23}) ({H}_{31}) + (\be%
_{3} - {a}_{3} - {n}_{12}) ({H}_{11})  \nonumber  \l{eq:03123} \\
&& \hspace*{1cm} - {\Omega}_{3} ({E}_{22} - {E}_{11}) + {\Omega}_{2} {E}%
_{23} - (2 {\omega}_{1} + {\Omega}_{1}) {E}_{31}  \nonumber \\
&& \hspace*{1cm} - \sfrac12 (3 {n}_{22} - {n}_{33} + {n}_{11}) {H}_{12} + ({a}%
_{2} - {n}_{31}) {H}_{23} + ({a}_{3} + {n}_{12}) {H}_{33} \\
\framebox[1.4cm]{[023]\,31} && ~~~ (\be_{0} + 3 \Th_{2}) ({E}_{12})
= (\be_{2} - 2 {a}_{2} - 2 {n}_{31}) ({H}_{23}) - (%
\be_{3} - {a}_{3} + {n}_{12}) ({H}_{22})  \nonumber  \l{eq:02331} \\
&& \hspace*{1cm} - {\Omega}_{3} ({E}_{22} - {E}_{11}) + ({\omega}_{1} - {%
\Omega}_{1}) {E}_{31} + {\Omega}_{2} {E}_{23}  \nonumber \\
&& \hspace*{1cm} - \sfrac12 (3 {n}_{11} + {n}_{22} - {n}_{33}) {H}_{12} - ({%
\dot{u}}_{1} + {a}_{1} + {n}_{23}) {H}_{31} - ({a}_{3} - {n}_{12}) {H}_{33}
\\
\framebox[1.4cm]{[023]\,01} && ~~~ (\be_{0} + 3 \Th_{2}) ({H}_{11})
= - (\be_{2} - {a}_{2} - {n}_{31}) ({E}_{31}) + ({\bf e%
}_{3} - {a}_{3} + {n}_{12}) ({E}_{12})  \nonumber  \l{eq:02301} \\
&& \hspace*{1cm} + \sfrac32 {n}_{11} {E}_{11} - \sfrac12 ({n}_{22}- {n}_{33}) (%
{E}_{22}- {E}_{33}) - 2 {n}_{23} {E}_{23} + 2 {\Omega}_{2} {H}_{31} - 2 {%
\Omega}_{3} {H}_{12} \\
\framebox[1.4cm]{[031]\,02} && ~~~ (\be_{0} + {\Th}) ({H}_{22}) = (%
\be_{1} + 2 {\dot{u}}_{1} - {a}_{1} + {n}_{23}) ({E}_{23}) - (\be%
_{3} - {a}_{3} - {n}_{12}) ({E}_{12})  \nonumber  \l{eq:03102} \\
&& \hspace*{1cm} + \sfrac32 {\sigma}_{11} {H}_{33} + \sfrac32 {n}_{22} {E}%
_{22} - \sfrac12 ({n}_{33}- {n}_{11}) ({E}_{33} - {E}_{11})  \nonumber \\
&& \hspace*{1cm} - 2 {n}_{31} {E}_{31} - (2 {\Omega}_{1} + \omega_{1}) {H}%
_{23} + 2 {\Omega}_{3} {H}_{12} \\
\framebox[1.4cm]{[012]\,03} && ~~~ (\be_{0} + {\Th}) ({H}_{33}) = - (%
\be_{1} + 2 {\dot{u}}_{1} - {a}_{1} - {n}_{23}) ({E}_{23}) + (\be%
_{2} - {a}_{2} + {n}_{31}) ({E}_{31})  \nonumber  \l{eq:01203} \\
&& \hspace*{1cm} + \sfrac32 {\sigma}_{11} {H}_{22} + \sfrac32 {n}_{33} {E}%
_{33} - \sfrac12 ({n}_{11}- {n}_{22}) ({E}_{11} - {E}_{22})  \nonumber \\
&& \hspace*{1cm} - 2 {n}_{12} {E}_{12} + (2 {\Omega}_{1} + {\omega_{1}}) {H}%
_{23} - 2 {\Omega}_{2} {H}_{31} \\
\framebox[1.4cm]{[012]\,02} && ~~~ (\be_{0} + {\Th} + \sfrac32 {\sigma}%
_{11}) ({H}_{23}) = - (\be_{1} + {\dot{u}}_{1} - {a}_{1} - {n}_{23}) ({E}%
_{22}) + (\be_{2} - 2 {a}_{2} + 2 {n}_{31}) ({E}_{12})  \nonumber
\l{eq:01202} \\
&& \hspace*{1cm} + {\omega}_{1} ({H}_{22} - {H}_{11}) - {\Omega}_{1} ({H}%
_{33} - {H}_{22}) + {\Omega}_{3} {H}_{31} - {\Omega}_{2} {H}_{12}
- \sfrac16 \be_{1}(\mu)  \nonumber
\\
&& \hspace*{1cm} + \sfrac12 (3 {n}_{33} - {n}_{11} + {n}_{22}) {E}_{23} - ({a}%
_{3} - {n}_{12}) {E}_{31} - ({a}_{1} + {n}_{23}) {E}_{11} + {\dot{u}}_{1} {E}%
_{33} \\
\framebox[1.4cm]{[031]\,03} && ~~~ (\be_{0} + {\Th} + \sfrac32 {\sigma}%
_{11}) ({H}_{23}) = (\be_{1} + {\dot{u}}_{1} - {a}_{1} + {n}_{23}) ({E}%
_{33}) - (\be_{3} - 2 {a}_{3} - 2 {n}_{12}) ({E}_{31})  \nonumber
\l{eq:03103} \\
&& \hspace*{1cm} - {\omega}_{1} ({H}_{33} - {H}_{11}) - {\Omega}_{1} ({H}%
_{33} - {H}_{22}) - {\Omega}_{2} {H}_{12} + {\Omega}_{3} {H}_{31}
+ \sfrac16 \be_{1}(\mu) \nonumber
\\
&& \hspace*{1cm} + \sfrac12 (3 {n}_{22} + {n}_{33} - {n}_{11}) {E}_{23} + ( {a%
}_{2} + {n}_{31}) {E}_{12} + ({a}_{1} - {n}_{23}) {E}_{11} - {\dot{u}}_{1} {E%
}_{22}  \\
\framebox[1.4cm]{[012]\,01} && ~~~ (\be_{0} + {\Th}) ({H}_{31}) = - (%
\be_{1} + {\dot{u}}_{1} - 2 {a}_{1} - 2 {n}_{23}) ({E}_{12}) + (\be%
_{2} - {a}_{2} + {n}_{31}) ({E}_{11})  \nonumber  \l{eq:01201} \\
&& \hspace*{1cm} - {\Omega}_{2} ({H}_{11} - {H}_{33}) - {\Omega}_{3} {H}%
_{23} + (2 {\omega}_{1} + {\Omega}_{1}) {H}_{12} + \sfrac16 \be%
_{2} (\mu) \nonumber \\
&& \hspace*{1cm} + \sfrac12 (3 {n}_{33} + {n}_{11} - {n}_{22}) {E}_{31} + ({a}%
_{3} + {n}_{12}) {E}_{23} + ({a}_{2} - {n}_{31}) {E}_{22}  \\
\framebox[1.4cm]{[023]\,03} && ~~~ (\be_{0} + 3 \Th_{2}) ({H}_{31})
= - (\be_{2} - {a}_{2} - {n}_{31}) ({E}_{33}) + ({\bf e%
}_{3} - 2 {a}_{3} + 2 {n}_{12}) ({E}_{23})  \nonumber  \l{eq:02303} \\
&& \hspace*{1cm} - {\Omega}_{2} ({H}_{11} - {H}_{33}) - {\Omega}_{3} H_{23}
- ({\omega}_{1} - {\Omega}_{1}) {H}_{12} - \sfrac16
\be_{2}(\mu) \nonumber \\
&& \hspace*{1cm} + \sfrac12 (3 {n}_{11} - {n}_{22} + {n}_{33}) {E}_{31} - (\ud%
1 + {a}_{1} - {n}_{23}) {E}_{12} - ({a}_{2} + {n}_{31}) {E}_{22}  \\
\framebox[1.4cm]{[031]\,01} && ~~~ (\be_{0} + {\Th} ) ({H}_{12}) = (%
\be_{1} + {\dot{u}}_{1} - 2 {a}_{1} + 2 {n}_{23}) ({E}_{31}) - (\be%
_{3} - {a}_{3} - {n}_{12}) ({E}_{11})  \nonumber  \l{eq:03101} \\
&& \hspace*{1cm} - {\Omega}_{3} ({H}_{22} - {H}_{11}) + {\Omega}_{2} {H}%
_{23} - (2 {\omega}_{1} + {\Omega}_{1}) {H}_{31} - \sfrac16 \be%
_{3}(\mu) \nonumber \\
&& \hspace*{1cm} + \sfrac12 (3 {n}_{22} - {n}_{33} + {n}_{11}) {E}_{12} - ({a}%
_{2} - {n}_{31}) {E}_{23} - ({a}_{3} + {n}_{12}) {E}_{33}  \\
\framebox[1.4cm]{[023]\,02} && ~~~ (\be_{0} + 3 \Th_{2}) ({H}_{12})
= - (\be_{2} - 2 {a}_{2} - 2 {n}_{31}) ({E}_{23}) + (%
\be_{3} - {a}_{3} + {n}_{12}) ({E}_{22})  \nonumber  \l{eq:02302} \\
&& \hspace*{1cm} - {\Omega}_{3} ({H}_{22} - {H}_{11}) + {\Omega}_{2} {H}%
_{23} + ({\omega}_{1} - {\Omega}_{1}) {H}_{31} + \sfrac16 \be_{3}(\mu)
 \nonumber \\
&& \hspace*{1cm} + \sfrac12 (3 {n}_{11} + {n}_{22} - {n}_{33}) {E}_{12} + ({%
\dot{u}}_{1} + {a}_{1} + {n}_{23}) {E}_{31} + ({a}_{3} - {n}_{12}) {E}_{33}
\ .
\eea
%

\subsection{Constraint equations}
\subsubsection{Momentum conservation constraints}
We contract the second Bianchi identity (\r{eq:Bianchi2}) twice and
find that for PLRS cosmologies according to Definition \r{def:1}
\bea
\l{eq:1b;b}
\be_{1}(p) + (\mu+p)\,\dot{u}_{1} & = & 0 \\
\l{eq:2b;b}
\be_{2}(p) & = & 0 \\
\l{eq:3b;b}
\be_{3}(p) & = & 0 \ .
\eea
%

\subsubsection{Fluid vorticity, spatial commutation function and
fluid shear constraints}
%
\bea
\l{eq:R0202R0303}
R_{0202} - R_{0303} & = & (E_{22}-E_{33}) = -\,2\,n_{23}\,
\dot{u}_{1} \\
\l{eq:R0203Jac0023}
R_{0203} - \left({}^{~\,0}_{0 2 3}\right)
& = & E_{23} = \sfrac{1}{2}\,(n_{22}-n_{33})\,\dot{u}_{1} \\
\l{eq:R0102}
R_{0102} & = & E_{12} = \be_{2}(\dot{u}_{1})
+ \Omega_{2}\,\omega_{1} + \sfrac{3}{2}\,\Omega_{3}\,\sigma_{11} \\
\l{eq:R0103}
R_{0103} & = & E_{31} = \be_{3}(\dot{u}_{1})
+ \Omega_{3}\,\omega_{1} - \sfrac{3}{2}\,\Omega_{2}\,\sigma_{11} \\
\l{eq:Jac0012}
\left({}^{~\,0}_{0 1 2 }\right)
& & 0 = \sfrac{1}{2}\,\be_{2}(\dot{u}_{1})
- \sfrac{1}{2}\,(a_{2}-n_{31})\,\dot{u}_{1}
+ \Omega_{2}\,\omega_{1} \\
\l{eq:Jac0013}
\left({}^{~\,0}_{0 1 3}\right)
& & 0 = \sfrac{1}{2}\,\be_{3}(\dot{u}_{1})
- \sfrac{1}{2}\,(a_{3}+n_{12})\,\dot{u}_{1}
+ \Omega_{3}\,\omega_{1} \\
\l{eq:Jac0123}
\left({}^{~\,0}_{1 2 3}\right)
& & 0 = (\be_{1}-\dot{u}_{1}-2\,a_{1})\,({\omega}_{1}) \\
\l{eq:R0123}
R_{0123} & = & \sfrac{3}{2}\,n_{11}\,\sigma_{11}
+ 2\,(\ud1+a_{1})\,\omega_{1}
\ = \ -\,H_{11} \\
\l{eq:R0231}
R_{0231} & = & -\,(\be_{1}-a_{1}+n_{23})\,\omega_{1}
- \sfrac34\,n_{11}\,\sigma_{11}
- \sfrac34\,(n_{22}-n_{33})\,\sigma_{11}
\ = \ -\,H_{22} \\
\l{eq:R0312}
R_{0312} & = & -\,(\be_{1}-a_{1}-n_{23})\,\omega_{1}
- \sfrac34\,n_{11}\,\sigma_{11}
+ \sfrac34\,(n_{22}-n_{33})\,\sigma_{11}
\ = \ -\,H_{33} \\
\l{eq:R0331}
R_{0331} & = & \be_{1}(\Th_{2})
+ \sfrac{3}{2}\,(a_{1}-n_{23})\,\sigma_{11}
+ \sfrac{1}{2}\,(n_{11}+n_{22}-n_{33})\,\omega_{1}
\ = \ - \,H_{23} \\
\l{eq:R0212}
R_{0212} & = & -\,\be_{1}(\Th_{2})
- \sfrac{3}{2}\,(a_{1}+n_{23})\,\sigma_{11}
+ \sfrac{1}{2}\,(n_{22}-n_{33}-n_{11})\,\omega_{1}
\ = \ - \,H_{23} \\
\l{eq:R0112}
R_{0112} & = & \be_{2}(\Th_{1})
- \sfrac{3}{2}\,(a_{2}-n_{31})\,\sigma_{11}
+ (a_{3}+n_{12})\,\omega_{1}
\ = \ -\,H_{31} \\
\l{eq:R0323}
R_{0323} & = & -\,\be_{2}(\Th_{2})
+ \be_{3}(\omega_{1})
\ = \ -\,H_{31} \\
\l{eq:R0131}
R_{0131} & = & - \,\be_{3}(\Th_{1})
+ \sfrac{3}{2}\,(a_{3}+n_{12})\,\sigma_{11}
+ (a_{2}-n_{31})\,\omega_{1}
\ = \ -\,H_{12} \\
\l{eq:R0223}
R_{0223} & = & \be_{3}(\Th_{2})
+ \be_{2}(\omega_{1})
\ = \ -\,H_{12} \\
\l{eq:Jac1123}
\left({}^{~\,1}_{1 2 3}\right)
& & \be_{1}(n_{11}) + \be_{2}(a_{3}+n_{12})
- \be_{3}(a_{2}-n_{31}) - 2\,a_{1}\,n_{11}
- 2\,a_{2}\,n_{12} - 2\,a_{3}\,n_{31}  \nonumber \\
& & = -\,2\,\Th_{1}\,\omega_{1} \\
\l{eq:R3112}
R_{3112} & = & -\,\be_{3}(a_{2}-n_{31})
+ \sfrac12\,\be_{1}(n_{11}+ n_{22}-n_{33})
- n_{11}\,(a_{1} - n_{23}) \nonumber \\
& & - \ a_{1}\,(n_{22}-n_{33}) - n_{23}\,(n_{22}+n_{33})
- 2\,n_{31}\,(a_{3}+n_{12}) + \Th_{1}\,\omega_{1}  \nonumber \\
& = & -\,E_{23} \\
\l{eq:Jac2123}
\left({}^{~\,2}_{1 2 3}\right)
& & \be_{2}(n_{22}) - \be_{1}(a_{3}-n_{12})
+ \be_{3}(a_{1}+n_{23}) - 2\,a_{1}\,n_{12} - 2\,a_{2}\,n_{22}
- 2\,a_{3}\,n_{23}  \nonumber \\
& & = -\,2\,\Omega_{3}\,\omega_{1} \\
\l{eq:R2312}
R_{2312} & = & -\,\be_{3}(a_{1}+n_{23})
- \sfrac12\,\be_{2}(n_{11}+n_{22}-n_{33})
+ n_{11}\,(a_{2}-n_{31}) \nonumber \\
& & + \ (a_{2}+n_{31})\,(n_{22}-n_{33})
+ 2\,n_{23}\,(a_{3}-n_{12}) - 2\,\Omega_{3}\,\omega_{1}
\nonumber \\
& = & -\,E_{31} \\
\l{eq:Jac3123}
\left({}^{~\,3}_{1 2 3}\right)
& & \be_{3}(n_{33}) - \be_{2}(a_{1}-n_{23})
+ \be_{1}(a_{2}+n_{31}) - 2\,a_{1}\,n_{31} - 2\,a_{2}\,n_{23}
- 2\,a_{3}\,n_{33}  \nonumber \\
& & = 2\,\Omega_{2}\,\omega_{1} \\
\l{eq:R2331}
R_{2331} & = & -\,\be_{2}(a_{1}-n_{23})
- \sfrac12\,\be_{3}(n_{22}-n_{33}-n_{11})
- n_{11}\,(a_{3}+n_{12}) \nonumber \\
& & + \ (a_{3}-n_{12})\,(n_{22}- n_{33})
- 2\,n_{23}\,(a_{2}+n_{31}) - 2\,\Omega_{2}\,\omega_{1}
\nonumber \\
& = & -\,E_{12} \\
\l{eq:R2323}
R_{2323} & = & \be_{2}(a_{2}+n_{31}) + \be_{3}(a_{3}-n_{12})
+  \Th_{2}^{2} + \omega_{1}^{2}
\nonumber \\
& & - \ (a_{1}+n_{23})\,(a_{1}-n_{23})
- (a_{2}+n_{31})^{2} - (a_{3}-n_{12})^{2}  \nonumber \\
& & - \ \sfrac34\,n_{11}^{2} + \sfrac12\,n_{11}\,(n_{22}+n_{33})
+ \sfrac14\,(n_{22}-n_{33})^{2} - 2\,\Omega_{1}\,\omega_{1}
\nonumber \\
& = & -\,E_{11} + \sfrac13\,\mu \\
\l{eq:R3131}
R_{3131} & = & \be_{3}(a_{3}+n_{12})
+ \be_{1}(a_{1}-n_{23}) +  \Th_{2}\,\Th_{1}  \nonumber \\
& & - \ (a_{2}+n_{31})\,(a_{2}-n_{31})
- (a_{3}+n_{12})^{2} - (a_{1}-n_{23})^{2}  \nonumber \\
& & + \ \sfrac14\,n_{11}^{2} + \sfrac14\,(n_{22}-n_{33})\,
(2\,n_{11}-3\,n_{22}-n_{33}) \nonumber \\
& = & -\,E_{22} + \sfrac13 \mu \\
\l{eq:R1212}
R_{1212} & = & \be_{1}(a_{1}+n_{23})
+ \be_{2}(a_{2}-n_{31}) +  \Th_{2}\,\Th_{1} \nonumber \\
& & - \ (a_{3}+n_{12})\,(a_{3}-n_{12})
- (a_{1}+n_{23})^{2} - (a_{2}-n_{31})^{2}  \nonumber \\
& & + \ \sfrac14\,n_{11}^{2} - \sfrac14\,(n_{22}-n_{33})\,
(2\,n_{11}-3\,n_{33}-n_{22}) \nonumber \\
& = & -\,E_{33} + \sfrac13\,\mu \ .
\eea
%

\subsubsection{Generalised Friedmann equation}
This is obtained by writing out ${R^{\alpha \beta}}_{\alpha
\beta}$ (which corresponds to summing (\r{eq:R2323}),
(\r{eq:R3131}) and (\r{eq:R1212})).
\bea
\l{eq:Friedmann}
& & 4\,\be_{1}(a_{1}) + 4\,\be_{2}(a_{2}) + 4\,\be_{3}(a_{3})
- 6\,a_{1}^{2} - 6\,a_{2}^{2} - 6\,a_{3}^{2} \nonumber \\ 
& & \hsh - \ 2\,n_{23}^{2} - 2\,n_{31}^{2} - 2\,n_{12}^{2}
- \sfrac12\,n_{11}^{2} - \sfrac12\,(n_{22}-n_{33})^{2}
+  n_{11}\,(n_{22}+n_{33}) \nonumber \\
& & \hsh + \ \sfrac23\,\Th^{2} - \sfrac32\,\sigma_{11}^{2}
+ 2\,\omega_{1}^{2} - 4\,\Omega_{1}\,\omega_{1}
=  2\,\mu \ .
\eea
%

\subsubsection{Constraints on ${\bf E}$ and ${\bf H}$}
%
\bea
\framebox[1.4cm]{[123]\,23} && ~~~ (\be_{1} - 3 {a}_{1}) ({E}_{11}) + (%
\be_{2} - 3 {a}_{2} + {n}_{31}) ({E}_{12}) + (\be_{3} - 3 {a}_{3} - {%
n}_{12}) ({E}_{31}) =  \nonumber  \l{eq:12323} \\
&& ~~~ \sfrac13 \be_{1} (\mu) + ({n}_{22} - {n}_{33}) {E}_{23} - ({E}%
_{22} - {E}_{33}) {n}_{23} + 3 {\omega}_{1} {H}_{11} \\
\framebox[1.4cm]{[123]\,31} && ~~~ (\be_{2} - 3 {a}_{2}) ({E}_{22}) + (%
\be_{3} - 3 {a}_{3} + {n}_{12}) ({E}_{23}) + (\be_{1} - 3 {a}_{1} - {%
n}_{23}) ({E}_{12}) =  \nonumber  \l{eq:12331} \\
&& ~~~ \sfrac13 \be_{2} (\mu) + ({n}_{33} - {n}_{11}) {E}_{31} - ({E}%
_{33} - {E}_{11}) {n}_{31} + 3 {\omega}_{1} {H}_{12} - \sfrac32 {\sigma}_{11}
{H}_{31} \\
\framebox[1.4cm]{[123]\,12} && ~~~ (\be_{3} - 3 {a}_{3}) ({E}_{33}) + (%
\be_{1} - 3 {a}_{1} + {n}_{23}) ({E}_{31}) + (\be_{2} - 3 {a}_{2} - {%
n}_{31}) ({E}_{23}) =  \nonumber  \l{eq:12312} \\
&& ~~~ \sfrac13 \be_{3} (\mu) + ({n}_{11} - {n}_{22}) {E}_{12} - ({E}%
_{11} - {E}_{22}) {n}_{12} + 3 {\omega}_{1} {H}_{31} + \sfrac32 {\sigma}_{11}
{H}_{12} \\
\framebox[1.4cm]{[123]\,01} && ~~~ (\be_{1} - 3 {a}_{1}) ({H}_{11}) + (%
\be_{2} - 3 {a}_{2} + {n}_{31}) ({H}_{12}) + (\be_{3} - 3 {a}_{3} - {%
n}_{12}) ({H}_{31}) =  \nonumber  \l{eq:12301} \\
&& ~~~ - (\mu + p) {\omega}_{1} + ({n}_{22} - {n}_{33}) {H}_{23} - ({H}_{22}
- {H}_{33}) {n}_{23} - 3 {\omega}_{1} {E}_{11} \\
\framebox[1.4cm]{[123]\,02} && ~~~ (\be_{2} - 3 {a}_{2}) ({H}_{22}) + (%
\be_{3} - 3 {a}_{3} + {n}_{12}) ({H}_{23}) + (\be_{1} - 3 {a}_{1} - {%
n}_{23}) ({H}_{12}) =  \nonumber  \l{eq:12302} \\
&& ~~~ ({n}_{33} - {n}_{11}) {H}_{31} - ({H}_{33} - {H}_{11}) {n}_{31} - 3 {%
\omega}_{1} {E}_{12} + \sfrac32 {\sigma}_{11} {E}_{31} \\
\framebox[1.4cm]{[123]\,03} && ~~~ (\be_{3} - 3 {a}_{3}) ({H}_{33}) + (%
\be_{1} - 3 {a}_{1} + {n}_{23}) ({H}_{31}) + (\be_{2} - 3 {a}_{2} - {%
n}_{31}) ({H}_{23}) =  \nonumber  \l{eq:12303} \\
&& ~~~ ({n}_{11} - {n}_{22}) {H}_{12} - ({H}_{11} - {H}_{22}) {n}_{12} - 3 {%
\omega}_{1} {E}_{31} - \sfrac32 {\sigma}_{11} {E}_{12} \ .
\eea
%



\end{document}